\documentstyle[preprint,amsfonts,aps]{revtex}
\tightenlines
\newcommand{\rv}{{\bf r}}

\newcommand{\Rv}{{\bf R}}

\newcommand{\Mvec}{{\bf M}}

\newcommand{\lll}{{l_1l_2l}}
\newcommand{\rd}{r_{12}}

\newcommand{\mulm}{\mu_{lm}}

\newcommand\eqref[1]{Eq.~(\ref{#1})}
\newcommand\klref[1]{(\ref{#1})}
\newcommand\cref[1]{Ref.~\cite{#1}}
\newcommand\abbref[1]{Fig.~\ref{#1}}
\newcommand\eqpref[1]{Eq.~(\protect\ref{#1})}
\newcommand\abbpref[1]{Fig.~\protect\ref{#1}}
\newcommand\pref[1]{\protect\ref{#1}}
\renewcommand\Re{\,{\rm Re}}
\renewcommand\Im{\,{\rm Im}}
\newcommand{\nn}{\nonumber}
\newcommand\av[1]{\left\langle #1 \right\rangle}
\newcommand\ts{T^\ast}
\newcommand\ms{{m^\ast}}
\newcommand\rhos{\rho^\ast}
\newcommand\ls{L^\ast}

\begin{document}
\draft
\title{Inhomogeneous magnetization in dipolar ferromagnetic liquids}
\author{B. Groh and S. Dietrich}
\address{
  Fachbereich Physik, Bergische Universit\"at Wuppertal, \\
 D--42097 Wuppertal, Federal Republic of Germany}

\maketitle

\begin{abstract}

At high densities fluids of strongly dipolar spherical particles
exhibit spontaneous long-ranged orientational order. Typically, due to
demagnetization effects induced by the long range of the dipolar
interactions, the magnetization structure is spatially inhomogeneous
and depends on the shape of the sample. We determine this structure
for a cubic sample by the free minimization of an appropriate
microscopic density functional using simulated annealing. We find a
vortex structure resembling four domains separated by four domain
walls whose thickness increases proportional to the system size
$L$. There are indications that for large $L$ the whole configuration
scales with the system size. Near the axis of the mainly planar
vortex structure the direction of the magnetization escapes into
the third dimension or, at higher temperatures, the absolute value of
the magnetization is strongly reduced. Thus the orientational order is
characterized by two point defects at the top and the bottom of the
sample, respectively. The equilibrium structure in an external field
and the transition to a homogeneous magnetization for strong fields
are analyzed, too.

\end{abstract}
\bigskip
\section{Introduction}

At high densities strongly dipolar fluids can exhibit an
orientationally ordered liquid phase characterized by a spontaneous
magnetization. This has been demonstrated for the first time by Monte
Carlo simulations of dipolar soft spheres by Wei and Patey
\cite{Patey1,Patey2} and has been confirmed by simulations of hard spheres
\cite{Weis1,Weis2,Grest2}  and of  Stockmayer fluids
\cite{Grest}. Theoretical approaches leading to the same conclusion
include a mean-field theory for a dipolar lattice gas \cite{Sano}, a
generalized van der Waals theory \cite{WZ}, and different kinds of
density-functional theory \cite{Patey3,Letter,Paper}.  At low
densities the typical configurations of dipolar fluids exhibit chain
formation \cite{Smit,WeisCh,LevesqueCh,Grest2} which may inhibit the phase transition from the isotropic
vapor phase to the isotropic liquid phase
\cite{McGrother,Sear,RoijLett,OsipovCh}.  If the dipole moments are
{\it electric} as in molecular liquids the orientationally ordered
phase exhibits a spontaneous polarization; in the case of {\it
magnetic} dipoles as in ferrofluids, i.e., colloidal suspensions of
permanently ferromagnetic particles, one speaks of ferromagnetic order
and a spontaneous magnetization. In this paper we adopt the magnetic
language, keeping in mind that completely analogous phenomena occur in
the electric case (as long as no free charge carriers are present in
the fluid).

In both  simulations and analytic theories the dipolar
forces must be treated carefully due to their long range which may
give rise to  effects depending on the shape of the sample. It turns out that for all
sample shapes with the exception of a long needle the equilibrium
configuration is inhomogeneous with a spatially varying magnetization
$\Mvec(\rv)$ \cite{WZComment,Domains}. This leads to a shape {\it independent}
free energy, as is expected on general grounds \cite{Griffiths}. The
highly non-trivial problem to determine explicitly the spatial
distribution of this inhomogeneous magnetization
for a given sample shape has not yet been 
solved satisfactorily. In  simulations usually a homogeneous magnetization is
enforced by using an infinitely permeable surrounding of the
periodically repeated simulation cells. If instead the sample is 
surrounded by vacuum the simulation cell splits into two domains with
opposing magnetizations \cite{Patey2}. But this structure is clearly induced
by the artificial periodic boundary conditions. We addressed this
problem recently \cite{Domains} for the experimentally more
relevant case of open boundaries and obtained the following general
characterization of the equilibrium configurations: On a macroscopic
scale the absolute value of $\Mvec(\rv)$ is constant, its divergence
is zero, and  at the surfaces the normal component vanishes. However,
these properties do not yet enable one to construct the configuration
for a given shape. For a cubic sample we  determined \cite{Domains} the most stable
configuration under the constraint of sharp domain boundaries. It
consists of four triangular domains with $90^\circ$ domain walls in
between. In contrast to solid ferromagnets the number of domains does
not increase with the system size, but remains as small as
possible. On the other hand the assumption of sharp domain boundaries
is not justified for liquid systems. Due to the lack of a lattice
anisotropy one rather expects very thick domain walls \cite{DeGennes}.

Therefore, in order to analyze the domain configuration in more detail in the
present work we have performed a numerical minimization of an
approximate microscopic density functional which has been used before
\cite{Letter,Paper,Domains} to describe the ferromagnetically ordered
fluid. We have made no a priori assumptions regarding the symmetry of
the domain structure and have minimized with respect to a large
number ($10^4$--$10^5$) of parameters which represent the local
magnetization at mesh points within the sample. This approach is
similar in spirit to the determination of the magnetization structure
of solid ferromagnetic particles of micrometer size in the framework
of micromagnetic theory \cite{Micromagnetics,Dunlop,Newell}. But there are two
main differences. First, we do not constrain the absolute value of the
magnetization, thus allowing for the formation of less ordered regions. Second, we work
on the microscopic scale determined by the particle diameter, which
enables us to examine in detail the overall structure as well as the
structure of domain walls and the cores of topological defects.

A brief account of some of our results has been published in
\cref{SimannLett}. Here we provide the important derivation of the functional
from the underlying model, discuss the implementation of the simulated
annealing, and give a  thorough analysis of the resulting
configurations. We also analyze the consequences of the necessary
discretization and of the finite sample size. Furthermore we discuss in
detail the orientational structure in
an external field, which is relevant for the experiments with the
recently discovered metallic liquid ferromagnets \cite{Platzek,Reske,Albrecht}.

\section{Model and theoretical approach}

\subsection{Density-functional theory for Stockmayer fluids}

We consider Stockmayer fluids consisting of spherical particles with
fixed embedded point dipoles which interact via pairwise dispersion
and dipolar forces. The interaction potential $w=w_{LJ}+w_{dip}$ is
the sum of the Lennard-Jones potential
\begin{equation}
  w_{LJ}(\rd)=4\epsilon\left(\left({\sigma\over \rd}\right)^{12}
        -\left({\sigma\over \rd}\right)^6\right)
\end{equation}
and the dipolar potential 
\begin{equation}
  w_{dip}(\rv_{12},\omega,\omega')={{m^2}\over{\rd^3}}\left({\bf 
   \hat m}(\omega)\cdot{\bf \hat m}(\omega')-3({\bf \hat m}(\omega)\cdot
     \hat\rv_{12})({\bf \hat m}(\omega')\cdot \hat\rv_{12})\right)
   \Theta(\rd-\sigma).
\end{equation}
$\rv_{12}=\rv-\rv'=\rd \hat \rv_{12}$ denotes the interparticle vector,
$\omega=(\theta,\phi)$ and $\omega'$ are the orientations of the
dipole moments at $\rv$ and $\rv'$, respectively, with an
absolute value $m$; hats indicate unit vectors. At short distances
the interaction is cut off by the Heaviside function $\Theta$.

In order to study the spatially inhomogeneous configurations within a
ferromagnetically ordered fluid we employ the density-functional
theory  introduced in \cref{Frodl1}
and for which we have worked out analytical results previously
\cite{Letter,Paper,Domains}.

The configurations of the fluid are described by the spatially
constant number density $\rho$ and
the normalized orientational distribution $\alpha(\rv,\omega)$ so that
the probability density for finding a particle at point $\rv$
with the orientation $\omega$ is $\hat\rho(\rv,\omega)=\rho\,\alpha(\rv,\omega)$. The free energy density
functional is given by 
\begin{equation} \label{Fstart}
  F(\rho,[\alpha(\rv,\omega)];T)=V f_{HS}(\rho,T)+F_{or}+F_{ex}+F_{H}.
\end{equation}
$V$ is the sample volume and $f_{HS}$ is the free energy density of the hard sphere reference
system characterized by an effective temperature dependent radius \cite{Frodl1}. The second term ($\beta=1/(k_B T)$)
\begin{equation} \label{Scont}
  F_{or}={\rho\over\beta} \int_V d^3r \int d\omega\,\alpha(\rv,\omega)
  \ln[4\pi \alpha(\rv,\omega)]
\end{equation}
takes into account the loss of entropy if the orientational
distribution is not isotropic, i.e., different from $1/(4\pi)$. Here
and in the following the  integrations over the angles $\omega$ are taken over the unit
sphere. Within the
low-density approximation the excess contribution due to the
long-ranged part of the interaction potential has the form
\cite{Frodl1}
\begin{equation} \label{Fexallg}
  F_{ex}=-{\rho\over{2\beta}} \int_V d^3r \int_V d^3r' \int d\omega
  d\omega' \alpha(\rv,\omega) \alpha(\rv',\omega') \Theta(\rd-\sigma)
  \tilde f(\rv_{12},\omega,\omega')
\end{equation}
with  the Mayer function 
\begin{equation}
  \tilde f(\rv_{12},\omega,\omega')=-1+\exp[-\beta
   w(\rv_{12},\omega,\omega')].
\end{equation}
This function can be expanded in terms of the rotational invariants ($C(\lll,m_1m_2m)$ are Clebsch-Gordan coefficients)
\begin{equation}
\Phi_\lll(\omega,\omega',\omega_{12})=\sum_{m_1,m_2,m}
C(\lll,m_1m_2m) Y_{l_1m_1}(\omega) Y_{l_2m_2}(\omega')
Y_{lm}^\ast(\omega_{12})
\end{equation}
with coefficients $\hat f$ depending only  on the  distance between
the particles:
\begin{equation} \label{fser}
  \tilde f(\rv_{12},\omega,\omega')=\sum_{\lll} \hat f_{\lll}(\rd)
  \Phi_\lll(\omega,\omega',\omega_{12}).
\end{equation}
Finally, the interaction energy with a homogeneous external field $H$ is 
\begin{equation} \label{FH}
  F_H=-\rho\, m H \int_V d^3r \int d\omega\,\alpha(\rv,\omega) \cos\gamma
\end{equation}
where $\gamma$ is the angle between $\omega$ and the  direction of the field.
We take into account the external potential due to the container
walls only summarily in that they provide the confinement of the fluid
to $V$ because we are interested in the {\it bulk} behavior of the
fluid. In this spirit we also do not consider spatial variations of
the number density $\rho(\rv)=\int d\omega\,\hat\rho(\rv,\omega)$ in the vicinity of the walls. Moreover, in
Eqs.~\klref{Fstart}--\klref{FH} we have assumed that $\rho(\rv)$
 is constant throughout the sample even if $\alpha(\rv,\omega)$
varies deep inside the sample. We expect that due to the small
compressibility of the fluid in the orientationally ordered phase possible variations of
$\rho(\rv)$, e.g. inside the domain walls, are only small and thus not
significant. 

In the next step the orientational distribution is expanded into spherical harmonics:
\begin{equation} \label{sphharm}
  \alpha(\rv,\omega)=\sum_{l=0}^\infty\sum_{m=-l}^l \mulm(\rv) Y_{lm}(\omega)
\end{equation}
with $\mu_{00}=1/\sqrt{4\pi}$ due to the normalization $\int d\omega\,
\alpha(\rv,\omega)=1$ and $\mulm^\ast=(-1)^m \mu_{l,-m}$ because
$\alpha$ is real. Inserting Eqs.~(\ref{fser}) and \klref{sphharm} into
\eqref{Fexallg} one finds for the excess free energy
\begin{eqnarray} \label{omintl}
 F_{ex} & = & -{{\rho^2}\over {2\beta}} \sum_{l_1,l_2,l} 
  \sum_{m_1,m_2,m} C(\lll,m_1m_2m) \nonumber \\
  & & \times \int\limits_V d^3r \int\limits_V d^3r' 
  \mu_{l_1m_1}(\rv) \mu_{l_2m_2}(\rv') \hat f_\lll(\rd) 
  Y_{lm}(\omega_{12}).
\end{eqnarray}
In order to make the free minimization of the density functional with
respect to the set $\{\mu_{lm}(\rv)\}$
numerically feasible we must refrain from determining the full
orientational distribution $\alpha(\rv,\omega)$ at each point $\rv$. Instead we focus on
the reduced information provided by the dimensionless local magnetization
\begin{equation} \label{mdef}
  \Mvec(\rv)=\int d\omega\,\alpha(\rv,\omega) \hat{\bf m}(\omega)
   =\left(\begin{array}{c}
    -\sqrt{2} \Re\,\mu_{11}(\rv) \\
     \sqrt{2} \Im\,\mu_{11}(\rv) \\
     \mu_{10}(\rv)  
    \end{array} \right)
   =\left(\begin{array}{c}
     M_1(\rv)\\
     M_2(\rv)\\
     M_3(\rv)
   \end{array} \right).
\end{equation}
(The actual magnetization is ${\cal M}(\rv)=\sqrt{4\pi/3}\,\rho m M(\rv)$.)
This implies that in \eqref{sphharm} only terms up to $l=1$ have to be taken
into account. Within this approximation those contributions to the excess
free energy which depend on the orientational order are given by
\begin{eqnarray}
  \Delta F_{ex}&=&F_{ex}^{(110)}+F_{ex}^{(112)} \nn \\
      & = & {{\rho^2}\over{2\beta}} {1\over\sqrt{12\pi}} \int_V d^3r
      \int_V d^3r'\,\Mvec(\rv)\cdot\Mvec(\rv') \Theta(\rd-\sigma) \hat
      f_{110}(\rd) \\
     & & +{{\rho^2}\over{2\beta}} \sqrt{5\over{24\pi}} \int_V d^3r
      \int_V d^3r'\,\Mvec(\rv) \hat{\bf T}(\hat{\rv}_{12}) \Mvec(\rv')
      \Theta(\rd-\sigma) \hat f_{112}(\rd) \nn
\end{eqnarray}
with the tensor $\hat T_{ij}(\hat\rv)=\delta_{ij}-3\hat r_i\hat
r_j$. We note that in contrast to \cref{Domains} we do not have to separate long- and short-ranged
contributions to $\hat f_{112}$ because within our numerical approach we are
always dealing with finite systems without carrying out the
thermodynamic limit.
Analytic expressions for the functions $\hat f_{11l}$ are provided by
Eqs.~(B33) and (B34) in \cref{Frodl1} from which one obtains the following
expansions in terms of $m^2$:
\begin{eqnarray} \label{f11lser}
  \hat f_{110}(r)&=& -{{8\pi}\over 25} \sqrt{{4\pi}\over 3} e^{-\beta
   w_{LJ}(r)} {{\beta^3 m^6}\over {r^9}} + O(m^{10}) \nn \\
  \hat f_{112}(r)&=&(4\pi)^{3/2}\sqrt{2\over 15}\left[ (1-e^{-\beta
   w_{LJ}(r)}) {{\beta m^2}\over {r^3}} + {6\over 25} e^{-\beta
  w_{LJ}(r)} {{\beta^3 m^6}\over {r^9}} + O(m^{10})\right].
\end{eqnarray}
To lowest order in $m^2$ the function $\hat f_{112}(r)$ contains the
characteristic dependence $\sim r^{-3}$ of the dipolar potential
whereas the function
$\hat f_{110}$ arises only due to the cubic and higher order terms in
the expansion of the Mayer function. (For the actual calculations we have
used the full expressions as given in \cref{Frodl1} instead of the 
expansions in \eqref{f11lser}.)

In the following we specificly analyze a cubic volume of side length
$L$ containing the fluid. The spatial integrations are discretized by
introducing a (simple cubic) lattice with lattice constant $a=L/N$ and
lattice vectors $\Rv=a (n_1,n_2,n_3)$, with $n_i \in
\{-(N-1)/2,\ldots,-1,0,1,\ldots,(N-1)/2\}$ ($N$ is even),
so that
\begin{equation} \label{Fexdis}
  \Delta F_{ex}={{\rho^2}\over {2\beta}} a^6
 \sum_\Rv \sum_{\Rv'} \sum_{i,j} M_i(\Rv) w_{ij}(\Rv-\Rv') M_j(\Rv').
\end{equation}
The interaction tensor $w_{ij}$ is given by
\begin{equation}
   w_{ij}(\Rv)=\sqrt{5\over{24\pi}} \Theta(R-\sigma) \left[\sqrt{2\over 5} \hat
   f_{110}(R)\delta_{ij}+\hat f_{112}(R)(\delta_{ij}-3 \hat R_i \hat
  R_j)\right], \qquad i,j=1,2,3,
\end{equation}
so that $w_{ij}(-\Rv)=w_{ij}(\Rv)$.
(The value of $\sigma/a$ is taken to be noninteger; otherwise there are
distances between lattice points which correspond to the discontinuity of the
Heaviside function.) An alternative formulation, which will prove to be
helpful later on, is
\begin{equation}
  \Delta F_{ex}={{\rho^2}\over{2\beta}} a^3 \sum_{\Rv} \sum_i H_i(\Rv)
   M_i(\Rv)
\end{equation}
with the local fields
\begin{equation} \label{Hlocdef}
   H_i(\Rv)=a^3 \sum_{\Rv'} \sum_j w_{ij}(\Rv-\Rv') M_j(\Rv').
\end{equation}

The entropic term given by \eqref{Scont} can be simplified using the fact that
by applying a suitable rotation the orientational distribution at a given point $\rv$
can be cast into the form
\begin{equation}
  \alpha(\rv,\omega)={1\over{4\pi}}+M(\rv) \sqrt{3\over{4\pi}}
  \cos\theta
\end{equation}
with $M(\rv)=|\Mvec(\rv)|$. Due to the angular integration in
\eqref{Scont} this
rotation does not alter the value of $F_{or}$ so that
\begin{equation}
  F_{or}={\rho\over\beta} \int_V d^3r \int d\omega
   \left({1\over{4\pi}}+M(\rv) \sqrt{3\over{4\pi}} \cos\theta\right)
   \ln\left(1+ M(\rv) \sqrt{12\pi}\cos\theta\right).
\end{equation}
The expansion of the logarithm yields in the discrete version
\begin{eqnarray} \label{Fordis}
  F_{or}&=& {\rho\over\beta} a^3 \sum_\Rv \sum_{n=1}^\infty 
  {{(\sqrt{12\pi}\, M(\Rv))^{2n}} \over {(2n-1) 2n (2n+1)}} \\
  &=& {\rho\over\beta} a^3 \sum_\Rv s(M(\Rv)). \nn
\end{eqnarray}
Finally, the term due to the external field, which is taken to point
into the $z$ direction, is
\begin{equation} \label{FHdis}
  F_{H}=-\sqrt{{4\pi}\over 3} \rho\, m H a^3 \sum_{\Rv} M_3(\Rv).
\end{equation}
Thus the free energy difference $F_{if}$ between the {\it f\/}erromagnetic and
the {\it i\/}sotropic phases is 
\begin{equation} \label{Fges}
   F_{if}=F_{or}+\Delta F_{ex}+F_H
\end{equation}
where the individual terms are given by Eqs.~\klref{Fexdis},
\klref{Fordis}, and \klref{FHdis}, respectively.

\subsection{Simulated Annealing} \label{Simann}

We have minimized the free energy in \eqref{Fges} with respect to the
magnetization configuration $\{\Mvec(\Rv)\}$ by using the simulated
annealing algorithm \cite{NumericalRecipes}. In the first step the
interaction tensor $w_{ij}$ is determined once for all relevant
distances between the lattice sites within $V=L^3$. In the second step
the local fields $H_i(\Rv)$ are
calculated for an initial choice of the configuration
$\Mvec^{(ini)}(\Rv)$ according to \eqref{Hlocdef}. In each following step a lattice site $\Rv_0$ is chosen
and a new magnetization $\Mvec'(\Rv_0)$ is proposed by 
changing each component $i=1,2,3$ by a random value $\Delta
M_i=M_i'(\Rv_0)-M_i(\Rv_0)$ between $-\kappa M(\Rv_0)$ and $+\kappa
M(\Rv_0)$ ($\kappa=0.1$ turned out to be a suitable choice). The resulting change in free energy is
\begin{equation}
  \Delta F_{if}={\rho\over\beta} a^3 \left[s(M'(\Rv_0))-s(M(\Rv_0)) +
   \sum_i H_i(\Rv_0) \Delta M_i\right] - \sqrt{{4\pi}\over 3}\rho\,m H a^3
  \Delta M_3.
\end{equation}
There is no  term quadratic in $\Delta M_i$ due to $w_{ij}(\Rv=0)=0$.
The proposed change of the configuration is accepted with certainty if
$\Delta F_{if}$ is negative and with a probability $\exp(-\Delta
F_{if}/(k_B T_s))$ if it is positive.  $T_s$ is the control temperature
of the annealing algorithm which is decreased slowly during the
minimization. In case of acceptance the new field $H_i'(\Rv)$ at each
site is calculated according to (see \eqref{Hlocdef})
\begin{equation}
  H_i'(\Rv)=H_i(\Rv)+a^3 \sum_j w_{ij}(\Rv-\Rv_0) \Delta M_j.
\end{equation}
The time required for this computational step, which is consuming most
of the CPU time, is of the order $O(N^3)$. The
advantage of using the local fields is that they need not be updated
if the proposed change is rejected, which happens quite often
especially at low control temperatures near the end of a run, while
the acceptance decision itself can be reached very fast. 

The control
temperature is lowered by a factor of $0.95$ after $3 N^3$ successful
changes or $15 N^3$ trials. The algorithm is terminated if no
successful step occurred during the $15 N^3$ trials. With the assumption that the
number of temperature steps is independent of $N$ the total computing
time should scale as $N^6$. Actually we found a scaling exponent
between 6 and 7. Typical CPU times for one run on a DEC alpha
workstation were 3.2 hours for $N=16$ and 25 hours for $N=22$. 

The minimizing configurations can be found starting from completely
random initial states, but this requires a relatively large initial
value of $T_s$ and therefore very long runs. For this reason we
started in almost all cases from a configuration which had been
obtained as a minimum for other parameter values. It is partly
randomized during the first phase of the algorithm by applying an
appropriate initial control temperature so that a large fraction of proposals
is accepted. By testing in some cases different starting
configurations we took care not to bias the final result by a
prejudiced initial guess.

\section{Results and discussion}

\subsection{Magnetization structure}

As our standard values of the thermodynamic parameters we chose
$\ms=\sqrt{m^2/\sigma^3\epsilon}=1.5$, $\ts=k_B T/\epsilon=2.25$, and
$\rhos=\rho\sigma^3=0.94$. This leads to a thermodynamic state which,
within the present density-functional theory approximation, lies deep
in the  ferromagnetic liquid phase. In view of the numerical
challenges described in Subcsec.~\ref{Simann} we have examined system sizes
$L^\ast=L/\sigma$ between 4.8 and 12 using lattices consisting of
$N=10,12,\ldots,24$ sites in each dimension. (Although we used an
ordinary workstation our maximum system size is even 30\%
larger than the system consisting of $22^3$ sites examined by Williams and Dunlop
\cite{Dunlop} on a supercomputer in 1989.)

The result of a minimization run is a three-component vector field
$\Mvec(\Rv)$ representing the magnetization structure within the cubic
volume. In order to visualize this field we display sections parallel
to the faces of the cube with the magnetization projected onto the
section plane. Since one component is lost due to this projection the
absolute value $|\Mvec(\Rv)|$ cannot be inferred from these
figures. We adopt a reference frame which has its origin in the center
of the cube. Figure~\ref{figschichtz}(a) shows a section perpendicular
to the $z$ axis at $z^\ast=z/\sigma=0.18\simeq 0$, i.e. close to the
center of the cube, for $L^\ast=7.2$ and
$N=20$. The overall picture is that of a vortex of closed magnetization lines
circulating around the $z$ axis. In this context it is interesting
to note that clusters of some ten to hundred dipolar particles also
exhibit a vortex structure at low temperatures
\cite{Singer1,Singer2}. This structure leads to $\text{div}\, {\bf H}=0$ for
the resulting magnetic field $\bf H$. A closer
look at the structure in
\abbref{figschichtz}(a) reveals that it may be described as composed of four domains
with an approximately constant magnetization separated by broad domain walls along the
diagonals of the square within which the direction of $\Mvec$ changes continuously.
This configuration resembles the triangular structure which
has been found to be the most stable structure under the
constraint of {\it sharp} domain walls \cite{Domains}. A similar
structure has also been found to be the most favorable one in cubic
magnetite particles just above the critical single-domain size, while
more complicated structures occur in this case for larger particles
\cite{Dunlop,Newell}. In a  section near the bottom
of the cube ($z^\ast=-3.06\simeq-L^\ast/2$, \abbref{figschichtz}(b))
the domain walls are slightly shifted off the diagonals into the
direction opposite to the circulation. The reverse situation is
found near the top. However, in contrast to the triangular
structure studied in \cref{Domains}, the magnetization is not confined to a
plane. As can be seen in the sections perpendicular to the $y$ axis
in \abbref{figschichty} there is a nonvanishing $z$ component, leading
to an ``escape into the third dimension'', which is particularly
pronounced near the vortex axis (Figs.~\ref{figschichty}(b) and
(c)). This mechanism avoids the formation of a topological defect
along the $z$ axis. This is in accordance with general considerations
showing that line
singularities are topologically unstable in a system of three-component spins in
three spatial dimensions \cite{Mermin}, which means that they can
always be removed by continuous {\it local} modifications. However, near the top
and the bottom face of the cube the $z$ component decreases in order
to avoid a large normal component at the surface which would produce
an unfavorable demagnetization field \cite{Domains}. Thus two
topologically stable point singularities near the centers of the
bottom and top surfaces are inevitable. A closer look
at the magnetization structure reveals a strong decrease of the
absolute value of $\Mvec$ in the core of these defects (see
\abbref{figmpdef}). The fact that the size of this less ordered core
also scales with the system size underlines the importance of
including the absolute value of the magnetization as a minimization
parameter. If instead the assumption of constant magnitude of $\Mvec$
were applied, as is usually done in the micromagnetic theory, the
defect could not be described correctly even on a mesoscopic scale.

Very similar structures have been found for all values of $L$ and
$N$. In \abbref{figauswmd} we compare the ``degree of escape'' for
different system sizes with fixed $N=20$. The cosine of the polar angle
of $\Mvec$ ($\cos\theta=M_3/M$) is plotted as a function of the
distance $r=\sqrt{x^2+y^2}$ from the center along the diagonals (i.e.,
the lines $x=\pm y$ with $z=\text{const}$) and the center lines
(i.e., the lines $x=0$ and $y=0$ with $z=\text{const}$). With the
exception of the smallest system the scaling property
\begin{equation} \label{scale}
  \Mvec(\rv/\sigma,L/\sigma)=\Mvec^{(0)}(\rv/L)
\end{equation}
is approximately fulfilled, so that one can surmise that this 
holds also  in the thermodynamic limit. (Since $\Mvec$ is
dimensionless, for a given thermodynamic state it can only depend on
$\rv/\sigma$ and $L/\sigma$.) The function $\Mvec^{(0)}$ represents
the global texture in the thermodynamic limit.
Near the edges of the cube,
i.e., near the corners in the
projection plane, there is an escape into the opposite direction, but
combined  with a strong decrease of the absolute value of the
magnetization (see \abbref{figmbetrmd}). Upon moving outwards from the
center $\cos\theta$ decreases faster along the diagonals than along
the center lines, indicating that there is not a circular but rather
a square symmetry.

In the inner part of the sample the absolute value $M(\Rv)=|\Mvec(\Rv)|$ (\abbref{figmbetrmd}) is approximately
constant and independent of $L$. It decreases near the surface and 
near the edges. This less ordered surface layer  thickens  relative
to the system size as $L$ decreases. Here, too, scaling with the
system size for large $L$ (\eqref{scale}) is compatible with the data. In the
smallest system we studied ($L^\ast=4.8$) $M$ decreases also near the center which
indicates a second mode for avoiding a line singularity besides the
``escape into the third dimension''. This mode appears also for the larger systems near the ferromagnetic-isotropic phase
transition and might be induced by the proximity of this transition
for $L^\ast=4.8$ (see Sec.~\ref{temp}). In \abbref{figmbetrz} we
present the average magnetization within each plane parallel to the
$xy$ plane $\langle M\rangle_{xy}=N^{-2} \sum_{n_1,n_2} M(n_1 a,n_2 a,z)$ as a
function of the height $z$. Again there is a clear decrease near the
surfaces while in the central region $\langle M \rangle_{xy}$ is
constant. The average $\langle M \rangle_{xy}$ attains its
thermodynamic limit $L\to\infty$ more rapidly than $\cos\theta$ or
$M(\Rv)$ (see Figs.~\ref{figauswmd} and \ref{figmbetrmd}). For small
$L$ the surface disordered region shrinks on the scale of $L$ but in
the thermodynamic limit it remains proportional to $L$. For large $L$
we find for the excess quantity $\int_{-L/2}^{L/2} dz \left[\langle M
\rangle_{xy}(0)-\langle M \rangle_{xy}(z)\right]/\langle M
\rangle_{xy}(0) \simeq 0.08 L$.

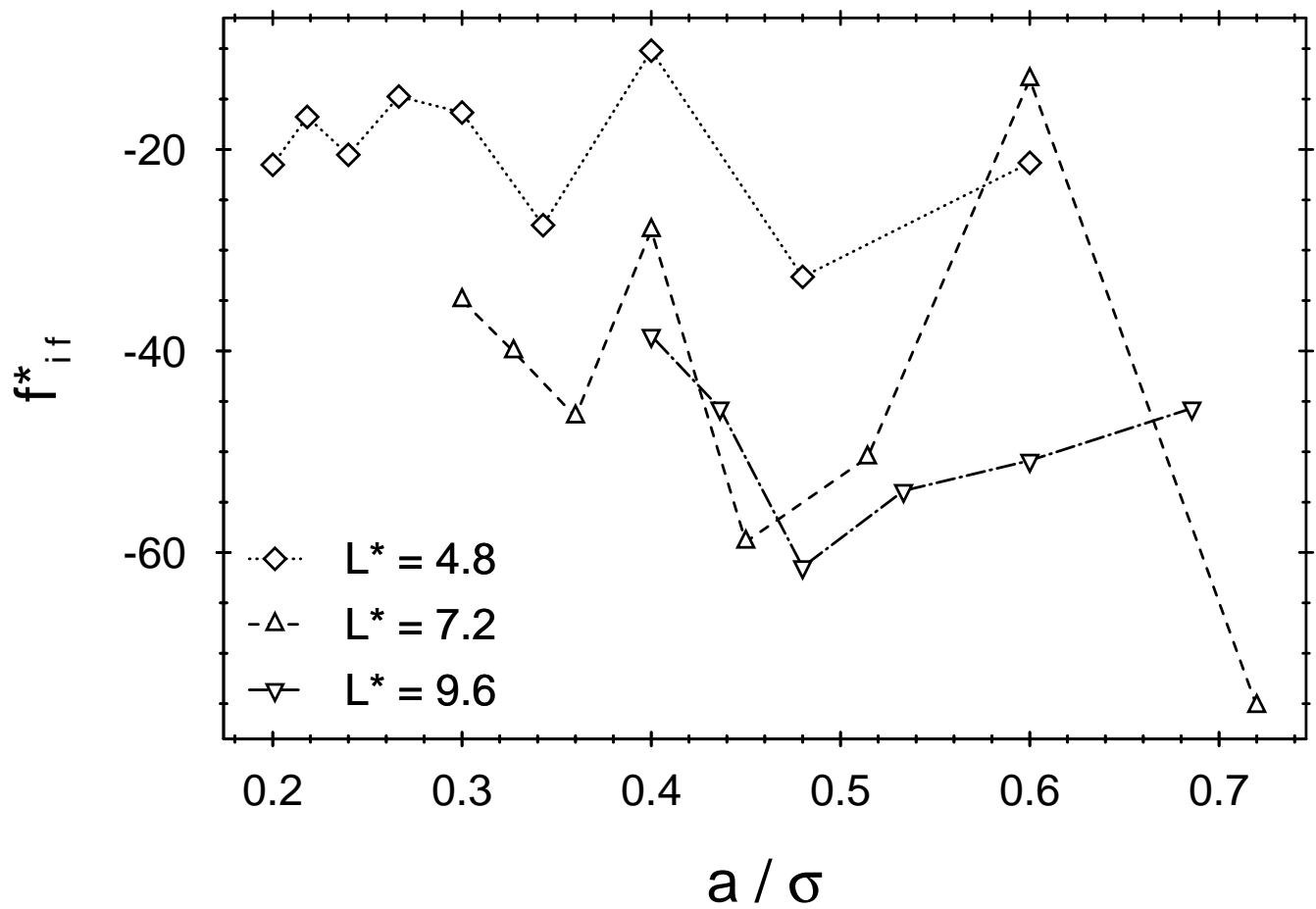
\begin{figure}
\caption{Reduced free energy difference between the isotropic and the
ferromagnetic phase
$f_{if}^\ast=({F_{if}}/{L^3})({\sigma^3}/\epsilon)$ as a
function of the lattice constant $a/\sigma=\ls/N$ of the discretization mesh for three
system sizes. The finite lattice constant induces
minima (maxima) near the points where the inverse $\sigma/a=N/\ls$ is
an integer (one half plus an integer). These oscillations
die out for $a\to 0$.}
\label{figenl}
\end{figure}

As shown in \abbref{figenl} the minimum value of the free energy density
$f^\ast_{if}=({F_{if}}/{ L^3}) ({\sigma^3}/\epsilon)$
exhibits a relatively strong, oscillatory dependence on the lattice
constant $a=L/N$ with a decreasing amplitude for increasing values of
$N$. This figure reveals that minima occur when 
$a/\sigma=L^\ast/N$ is close to the inverse of an integer, which indicates a strong discretization effect. For the
larger values of $L$ we could not reach the region where the
oscillations have died out, which inhibits a further finite-size
analysis. Similar oscillations were found in the spatially averaged value of
$M(\Rv)$ and for $L^\ast=4.8$ also in the degree of
the rotation out of the plane.  The precision to which
the minimum value of $f^\ast_{if}$ can be determined by simulated
annealing for a given set of parameters is much higher than the discretization
effect described above; from the dependence $f^\ast_{if}(T_s)$ we
estimate the minimization error to be of
the order of 0.01.

\subsection{Domain walls}

In solid ferromagnets the walls between adjacent domains
have a microscopic thickness  (i.e., it does not scale with the system
size) which is determined by the competition
between the exchange energy resulting from the spin coupling and the
anisotropy energy due to the lattice structure which causes easy axes
for the magnetization. Since there is no such lattice anisotropy in liquid ferromagnets
de Gennes and Pincus \cite{DeGennes} surmised that consequently there
are also no domain walls. Below we shall argue that in the case of
cubic samples as
considered here the thickness of the domain walls is proportional to
the system size and thus diverges in the thermodynamic
limit. Thus one is left with a question of terminology whether one still
speaks of walls, but certainly the behavior is qualitatively
different from that in solids.

In order to analyze the properties of the fluid domain walls in the
finite cube we consider the
behavior of the dimensionless magnetization $\Mvec$ along straight paths
normal to the wall. Along these paths $\Mvec$ changes continuously
between the magnetization directions of the adjacent domains. In the
present case these domains have the quasi-triangular structure
indicated in \abbref{figschichtz}(a) so that the orientational order
between neighboring domains differs
by an angle of $\pi/2$. Except near the vortex axis we
find a N\'eel type of wall, i.e., the magnetization vector
rotates mainly within the plane spanned by the asymptotic orientations deep
inside the adjacent domains; in \abbref{figschichtz}(a) this is
 the $xy$ plane. In contrast, in bulk solid
ferromagnets one usually observes Bloch walls with the magnetization vector
rotating out of plane on a cone around the wall normal \cite{Hubert};
only in thin solid films \cite{Hubert} and small particles
\cite{Dunlop} N\'eel walls
do occur, too. In \abbref{figmtxy} the relevant transversal component
$M_{t,xy}=(M_1+M_2)/\sqrt{2}$ is shown as a function of the normal
coordinate $n$ (compare \abbref{figschichtz}(a)). 

We define the wall thickness $\delta$  as the distance
between the points where the tangent to the curve at its inflection point
reaches the extreme values of $M_{t,xy}$ (see \abbref{figmtxy}). The dependence of $\delta$
on the distance $r$ of the wall normal from the center of the cube is
displayed in \abbref{figthick} (all paths lie in the plane
$z=-a/2$, i.e., close to the center of the cube).
 The wall thickness decreases near the edge of the cube where the
 normal paths hardly reach the region of homogeneous magnetization
(see \abbref{figschichtz}). The thickness increases near the center
$r=0$ due to
the ``escape'' region (see \abbref{figschichty}(c)). However, it is approximately constant in
the range $r/L=0.2\ldots0.35$ and there is no monotonic trend as
function of the
number of lattice sites. Its dependence on the system size $L$ is
analyzed in \abbref{figmidthick}. Here we selected the values of
$\delta$ at (or, due to the discrete lattice, close to) $r/L=1/2^{3/2}$ corresponding to half the
distance between the center and the edge. For each size $L$ we display the
results obtained for different values of $N$ which render an estimate
of the uncertainty caused by the finite lattice constant $a=L/N$. A slight  decrease of
$\delta/L$ with increasing $L$ indicates that the domains are getting
sharper. However, most probably the data can be extrapolated to a
finite limit of $\delta/L$ for $L\to\infty$ which would be in
accordance with the proposed scaling behavior in \eqref{scale}.

\subsection{Temperature dependence and critical point} \label{temp}

Starting from the standard value $\ts=k_B T/\epsilon=2.25$ used up to
here we have
increased the temperature at the fixed density $\rhos=0.94$ in order
to examine the structural changes upon
approaching the ferromagnetic-isotropic transition. We have chosen $N=16$
and studied the system sizes $L^\ast=4.8$ and $L^\ast=9.6$. The spatially
averaged absolute value of the magnetization decreases and finally
vanishes at a temperature $T_c(L)$ in accordance with the inherent
mean field approximation (see \abbref{figmagnt}). (We define the
finite-size critical temperature as the limiting temperature
above which
no configurations with a negative free energy difference
$F_{if}$ are found by the minimization algorithm.) As expected this
finite-size (strictly speaking quasi-) critical temperature is lower for smaller systems. From
Eq.~(7.10) in \cref{Paper} we infer $\ts_c(L\to\infty)=3.04$ for the
parameters used here. The evolution of the magnetization structure is
analyzed in Figs.~\ref{figauswmt} and \ref{figmbetrdt} for
$L^\ast=9.6$. The escape into the $z$ direction near the vortex axis
is strongly reduced at higher temperatures (\abbref{figauswmt}) while
 the absolute value of the magnetization in this
region decreases more rapidly  than for intermediate values of $r/L$
(\abbref{figmbetrdt}). Thus a column of less ordered fluid develops
around the vortex axis. These effects are even more pronounced in the
smaller system, as has been already suggested by Figs.~\ref{figauswmd} and
\ref{figmbetrmd}. Furthermore \abbref{figmbetrdt} demonstrates that
the  surface layer with reduced orientational order thickens,
which is a consequence of the increasing correlation length upon
approaching the phase transition. The domain wall thickness as defined
in the preceding subsection also increases slightly but this
change is smaller than the uncertainty caused by the finite lattice
constant.

\section{External field} \label{field}

Up to now all results refer to zero external field. If an external
field is applied the dipolar particles tend to align along the field
direction. The resulting transition from the inhomogeneous zero-field
configuration to the homogeneously magnetized state in the presence of strong fields can
also be examined within the present approach. 

The field destroys the equivalence of the three perpendicular
directions of the cubic axes. (One should keep in mind that below
$T_c(L)$ this equivalence is also {\it spontaneously}
broken in zero field.) We have applied the field normal to the
surfaces of the cube either parallel or perpendicular to the spontaneously chosen vortex axis of the
zero-field configuration, which was used as an initial guess for the
minimization algorithm. The most stable configurations were found when the
external field is parallel to the vortex axis. (In principle,
equivalent  but rotated configurations should be obtained if at the
beginning of the algorithm the field is applied perpendicular to this
axis. Since, however, only a medium value for the initial control
temperature was employed, for this latter choice of the initial guess the system could not reach the equilibrium
structure which in this case differs significantly from the initial
configuration.) The relative stability of the resulting configurations
can be judged on the basis of the corresponding value of the free energy. A typical result
is depicted in Figs.~\ref{figschichtzh} and
\ref{figschichtyh}. The section
parallel to the $xy$ plane, i.e., perpendicular to the external field, exhibits
smaller absolute values of the  magnetization components orthogonal to
the field direction
than for $H=0$ (\abbref{figschichtzh}), but the structure is very similar. On the other hand
there is a substantial increase of the magnetization component parallel
to the field as shown in the sections $y=\text{const}$ in
\abbref{figschichtyh}. Thus the overall structure is essentially preserved,
but at each point the magnetization is rotated into the field
direction by a certain amount. This demonstrates that this behavior,
which was anticipated  in
\cref{Domains},   corresponds indeed to
the equilibrium configuration. If the field is sufficiently strong the
vortex structure is lost at a critical field strength $H_c$. This
is shown in \abbref{figmagn} by the field dependence of the averaged
parallel component $\langle M_\parallel \rangle=N^{-3} \sum_\Rv
M_3(\Rv)$ and of the angular component $\langle M_\varphi \rangle=N^{-3}
\sum_\Rv {\bf e}_\varphi(\Rv)\cdot\Mvec(\Rv)$, where ${\bf
e}_\varphi(\Rv)=(R_2,-R_1,0)/|\Rv|$.  Below $H_c$ the parallel component
increases approximately linearly \cite{Domains} while
the angular component decreases and vanishes at $H_c$ (apparently
linearly in $H_c-H$). The absolute value of $\Mvec$ increases almost
homogeneously over the sample. A surface layer of lower magnetization
is preserved even in the presence of strong fields.
Above $H_c$ the increase of $\langle M_\parallel \rangle$ is only due
to the increase of $|\Mvec|$ and will saturate in the limit
$H\to\infty$. 

These results should be compared with those of a micromagnetic
calculation for isotropic {\it spheres} by Aharoni and Jakubovics
\cite{Aharoni}. They were intended as a first step in modelling
amorphous solid ferromagnets but due to the assumption of a vanishing
anisotropy constant they should also be of interest for liquid ferromagnets.
These authors find a vortex structure with the axis parallel to the
external field, too. In the center of the vortex the magnetization
points into the field direction, also for $H\to 0$. However, for the
cube this contribution to the net magnetization is compensated by the opposite orientation of the
magnetization near the edges (see \abbref{figauswmd}(b)). This implies
that for cubes
$\av{M_\parallel}\simeq 0$ for $H\to 0$, whereas for
spheres a spontaneous net magnetization $\av{M_\parallel}(H\to
0^+)=-\av{M_\parallel}(H\to 0^-)\neq 0$
remains in the limits $H\to 0^\pm$. This discontinuity at $H=0$ yields an
infinite zero field
susceptibility. Thus spontaneous liquid ferromagnetism should be easier to detect
by macroscopic magnetization measurements in spherical than in cubic
samples.

The same transition between an inhomogenous and a nearly homogeneous state
arises when the
external field is kept fixed and the temperature is varied. As shown in
\abbref{figmagnhfix} the angular magnetization component decreases
with increasing $T$ and vanishes at a critical temperature $T_c(H)$,
which is lower than $T_c(H=0)$ (compare \abbref{figmagnt}). The parallel
component is nearly constant below $T_c$ and decreases gradually above
$T_c$ where the sample is almost homogeneously magnetized. Our
previous work \cite{Domains} as well as \cref{Rayl} predict a kink in
the curve  $M_\parallel(T)$ and the constant value ${\cal
  M}(T<T_c)=H/(4\pi D)$ with the demagnetization factor $D$. In the
present units, using $D=1/3$ for the cube, this would mean
$M(T<T_c)=(3/4\pi)^{3/2} H/(\rho m)\simeq 0.0827$ for the parameters
used in \abbref{figmagnhfix}. Both features have also been observed in
{\it ellipsoidal} samples of a solid dipolar Ising ferromagnet
\cite{Frowein}. Obviously the present theory yields a higher plateau
value of the magnetization and a rounding of the kink at the phase
transition. One also observes that the angular component vanishes
linearly at the critical point, as in \abbref{figmagn}, but in
contrast to the simplified model used in \cref{Domains}. We surmise that
these differences are due to the fact that for a {\it cube} the 
assumption adopted in Refs.~\cite{Rayl} and \cite{Domains} of a
homogeneous magnetization $M_\parallel$ in
both phases is not fulfilled. Another possible source for this
dicrepancy could be the finite size of the systems we examined.

\section{Summary}

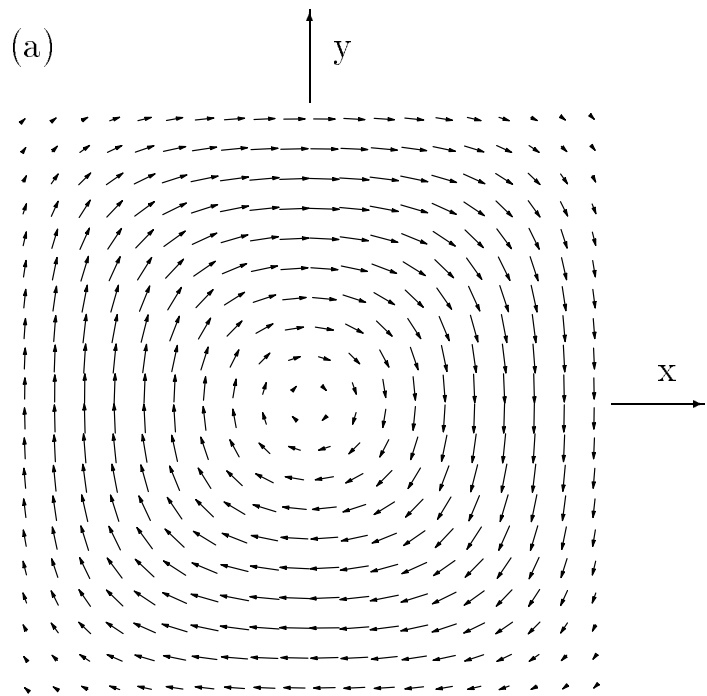
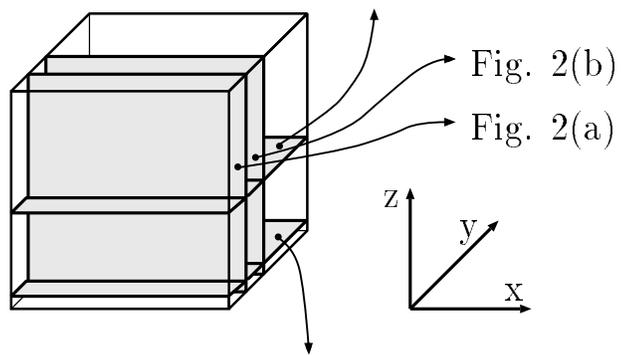
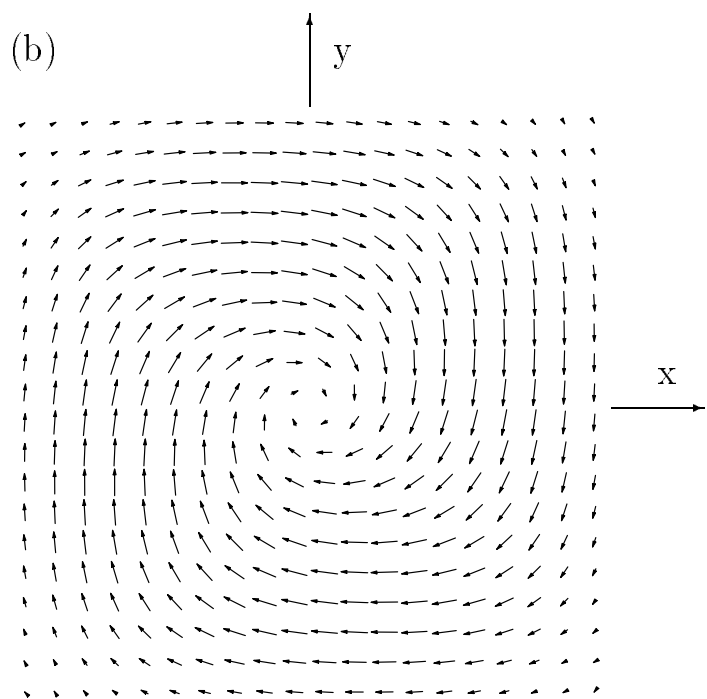
\begin{figure}
\caption{Sections perpendicular to the $z$ axis through the
magnetization structure of a ferromagnetic liquid in a cubic volume
(for $\ls=7.2$ and $N=20$); (a) section plane $z^\ast=0.18$ near the
midplane, (b) section plane $z^\ast=-3.06$ near the bottom. The arrows
represent projections of the local magnetization (at their midpoint)
onto the section plane.}
\label{figschichtz}
\end{figure}

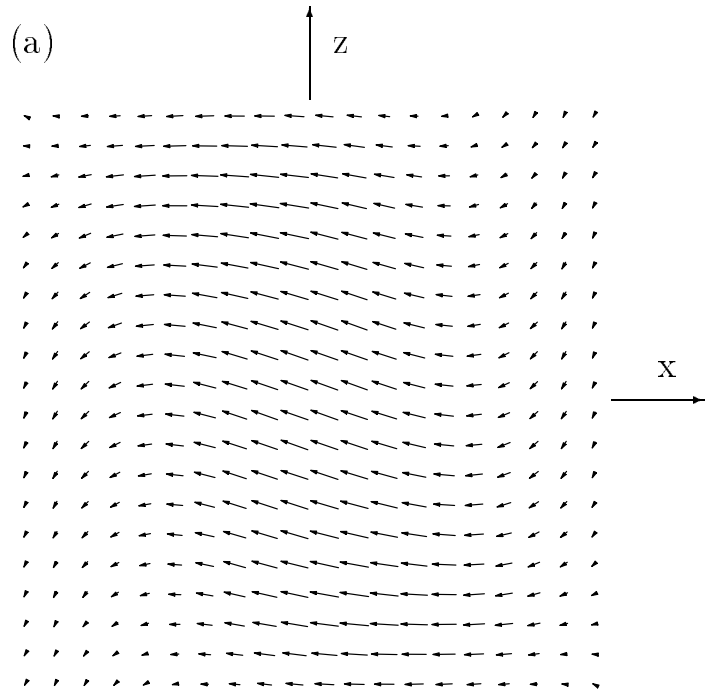
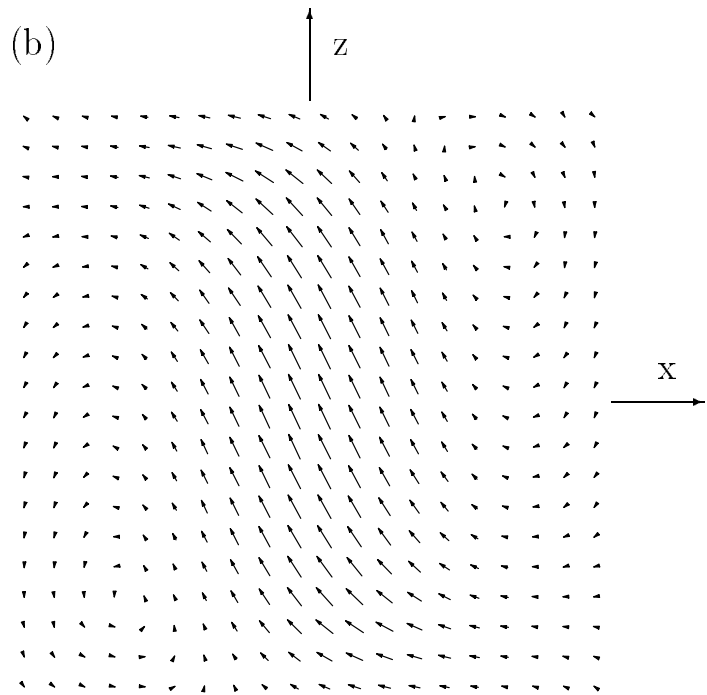
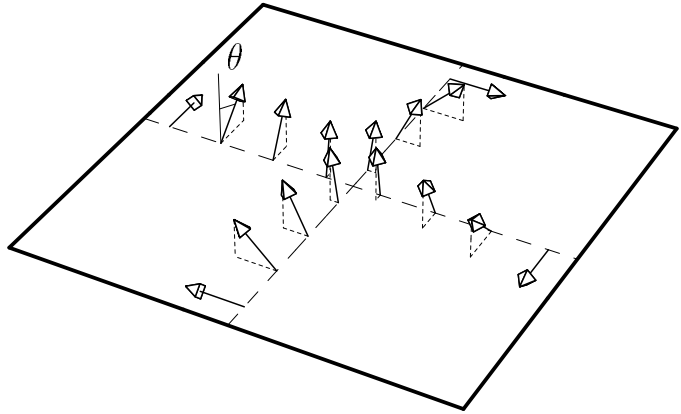
\begin{figure}
\caption{The same as in \abbpref{figschichtz}, but sections
perpendicular to the $y$ axis; (a) $y^\ast=-1.62$, (b)
$y^\ast=-0.54$. The schematic drawing in (c) demonstrates the
mechanism of the ``escape into the third dimension'' which avoids the
formation of a topological line defect along the $z$ axis. $\theta$ denotes the
polar angle of the magnetization. (The increase of $\theta$ to values
larger than $\pi/2$ close to the sample edges (see
\abbpref{figauswmd}(b)) is not shown.) }
\label{figschichty}
\end{figure}

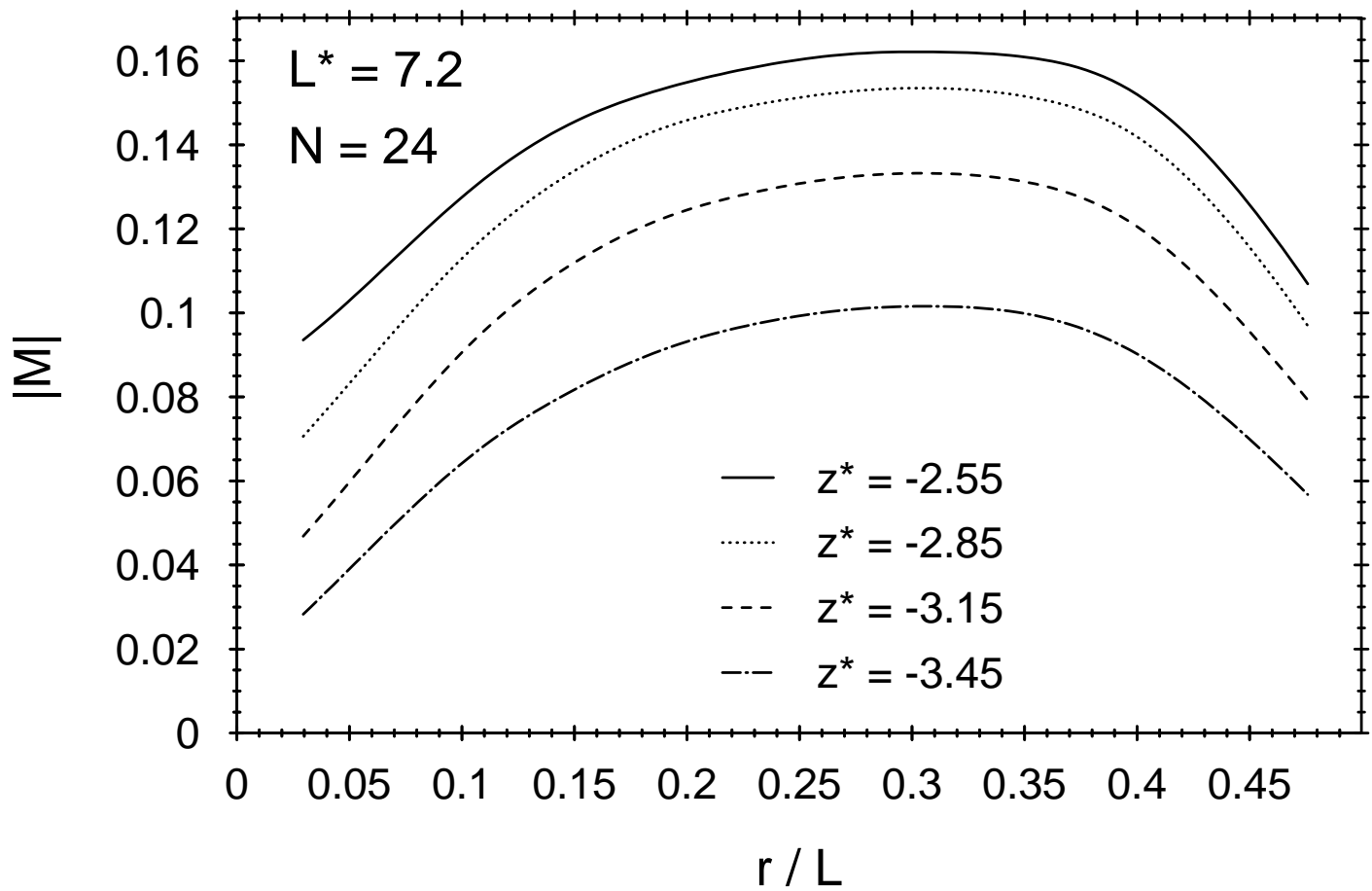
\begin{figure}
\caption{The absolute value of the dimensionless magnetization near
the point defect at the center of the bottom face of the cube (for
$\ls=7.2$ and $N=24$). $|\Mvec|$ is plotted as a function of the
distance from the $z$ axis along lines parallel to the $x$ or $y$ axis
for a series of fixed values of $z$ (compare the case $z\simeq 0$ in,
c.f., \abbpref{figmbetrmd}(a)). One finds a pronounced decrease of $|\Mvec|$ near
the core of the point defect.}
\label{figmpdef}
\end{figure}

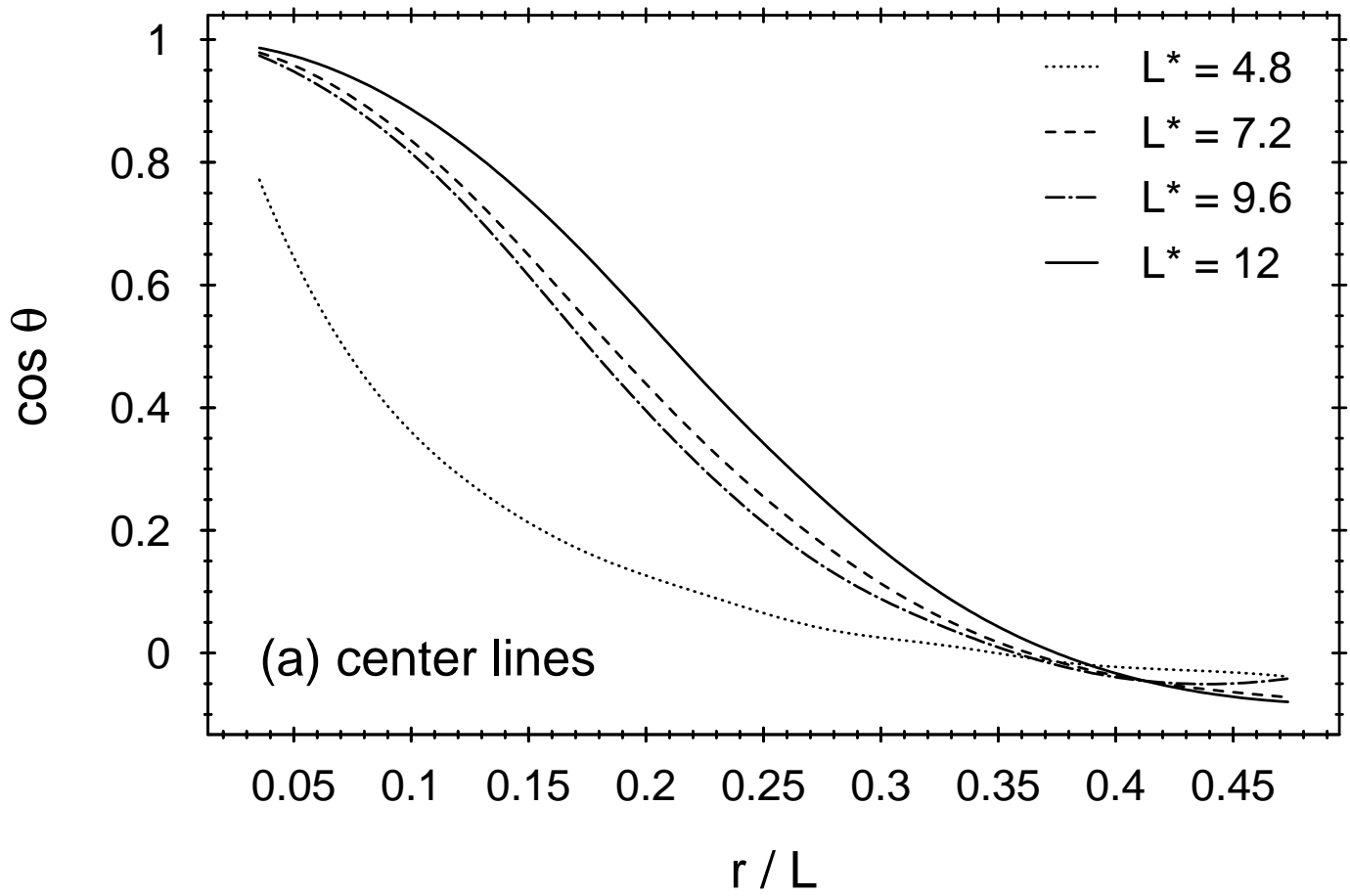
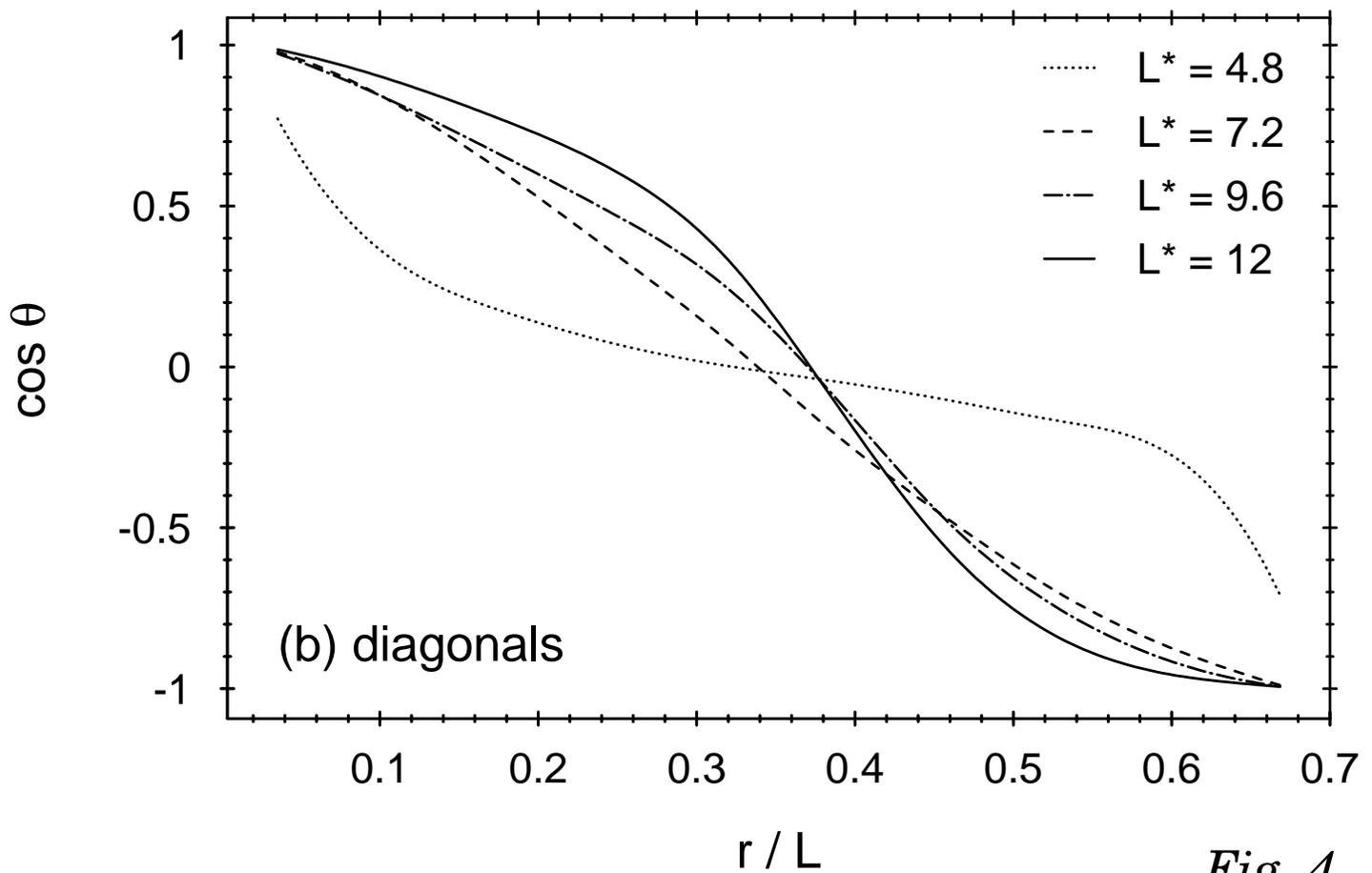
\begin{figure}
\caption{The cosine of the polar angle of the magnetization
($\cos\theta=M_3/M$) along (a) the center and (b) the diagonal lines
within a plane close to the midplane ($z^=-a/2$, with the lattice
constant $a=L/N$) for $N=20$ and
different system sizes $\ls$. $r=\protect\sqrt{x^2+y^2}$ is the
distance from the center of the cube; $r/L\leq 0.5$ for the center
lines (a) and $r/L\leq 1/\protect\sqrt{2}=0.707$ for the diagonals (b). 
The values of $\cos\theta$ are averaged over the four equivalent sites for each
value of $r/L$.
The
magnetization ``escapes'' into the positive $z$ direction near the
center ($\cos\theta\to 1$), lies in plane ($\cos\theta=0$) near the
sample surfaces (center lines), and turns into the negative $z$
direction ($\cos\theta=-1$) near the edges. Except for the smallest
system ($\ls=4.8$) the proposed scaling with the system size (see
\eqpref{scale}) is fulfilled approximately, i.e., in the limit $L\gg
\sigma$ two master curves evolve, one for the center lines and one for
the diagonals.}
\label{figauswmd}
\end{figure}

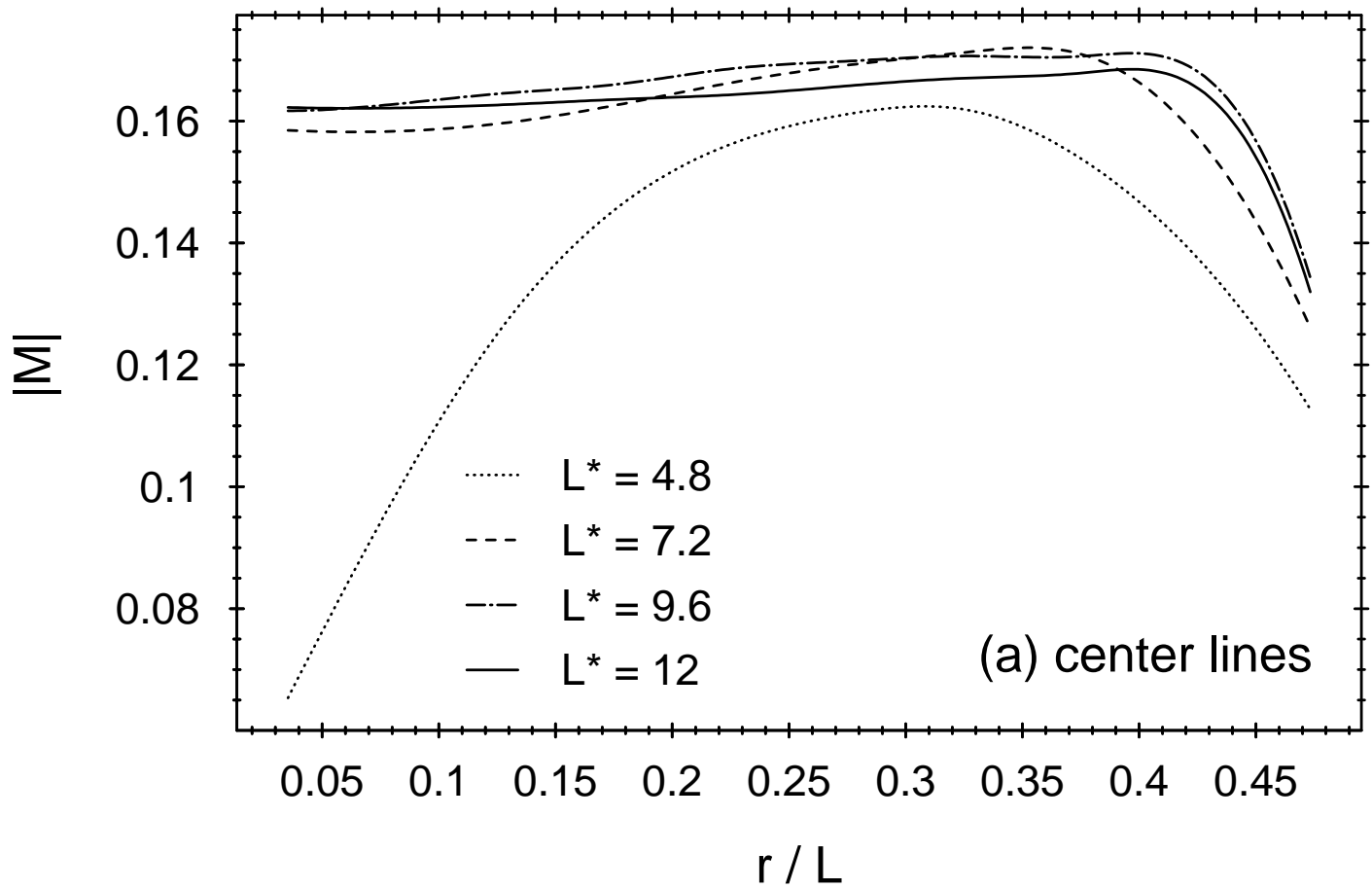
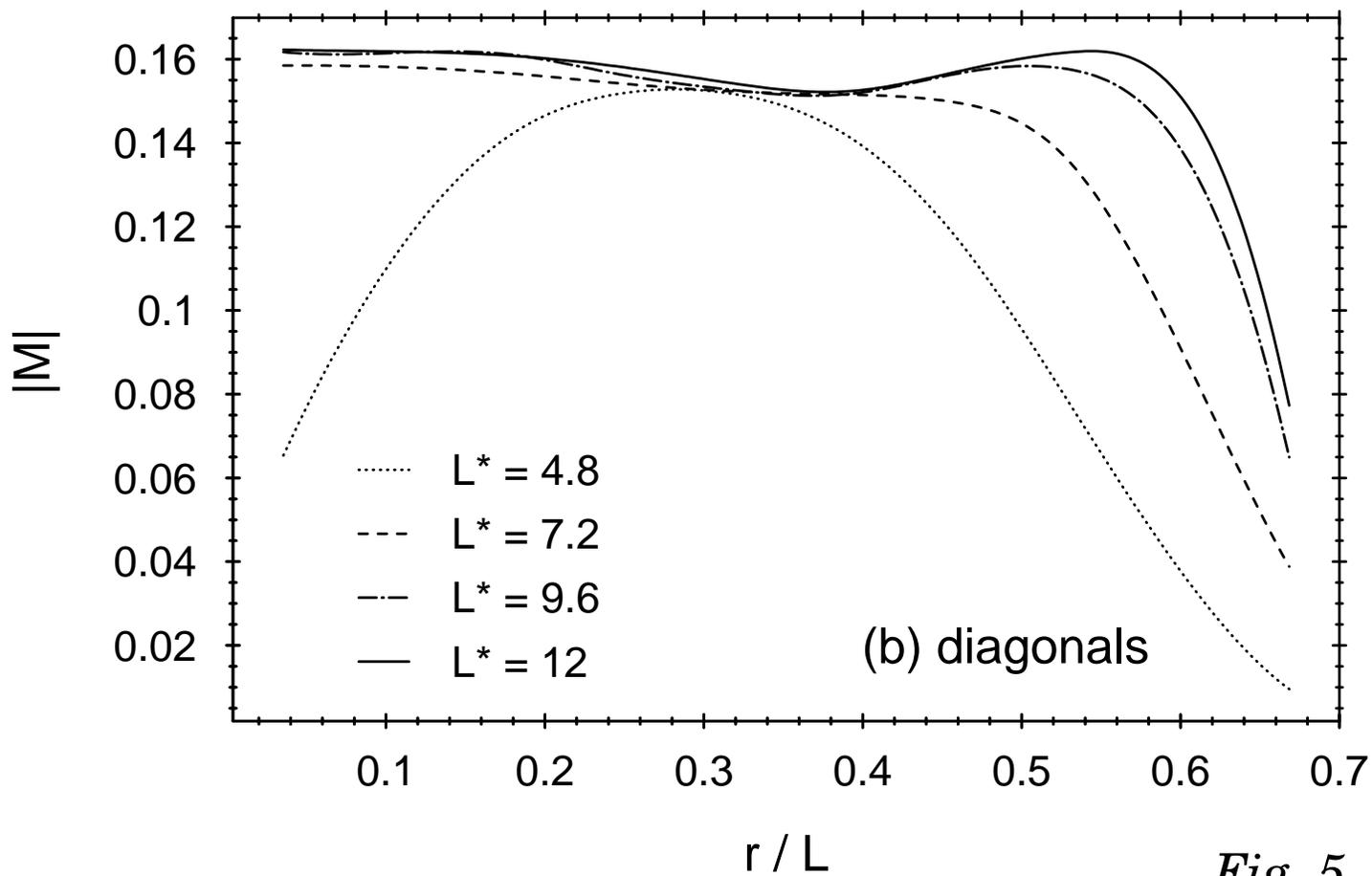
\begin{figure}
\caption{The absolute value $M$ of the dimensionless magnetization $\Mvec$
for the same parameters as in \abbpref{figauswmd} averaged over (a)
the four center lines and (b) the four diagonals. With the exception
of the smallest system $M$ is approximately constant in the bulk and
decreases near the surfaces. For $L\gg\sigma $ the behavior turns into
a single master curve consistent with the proposed scaling
(\eqpref{scale}).}
\label{figmbetrmd}
\end{figure}

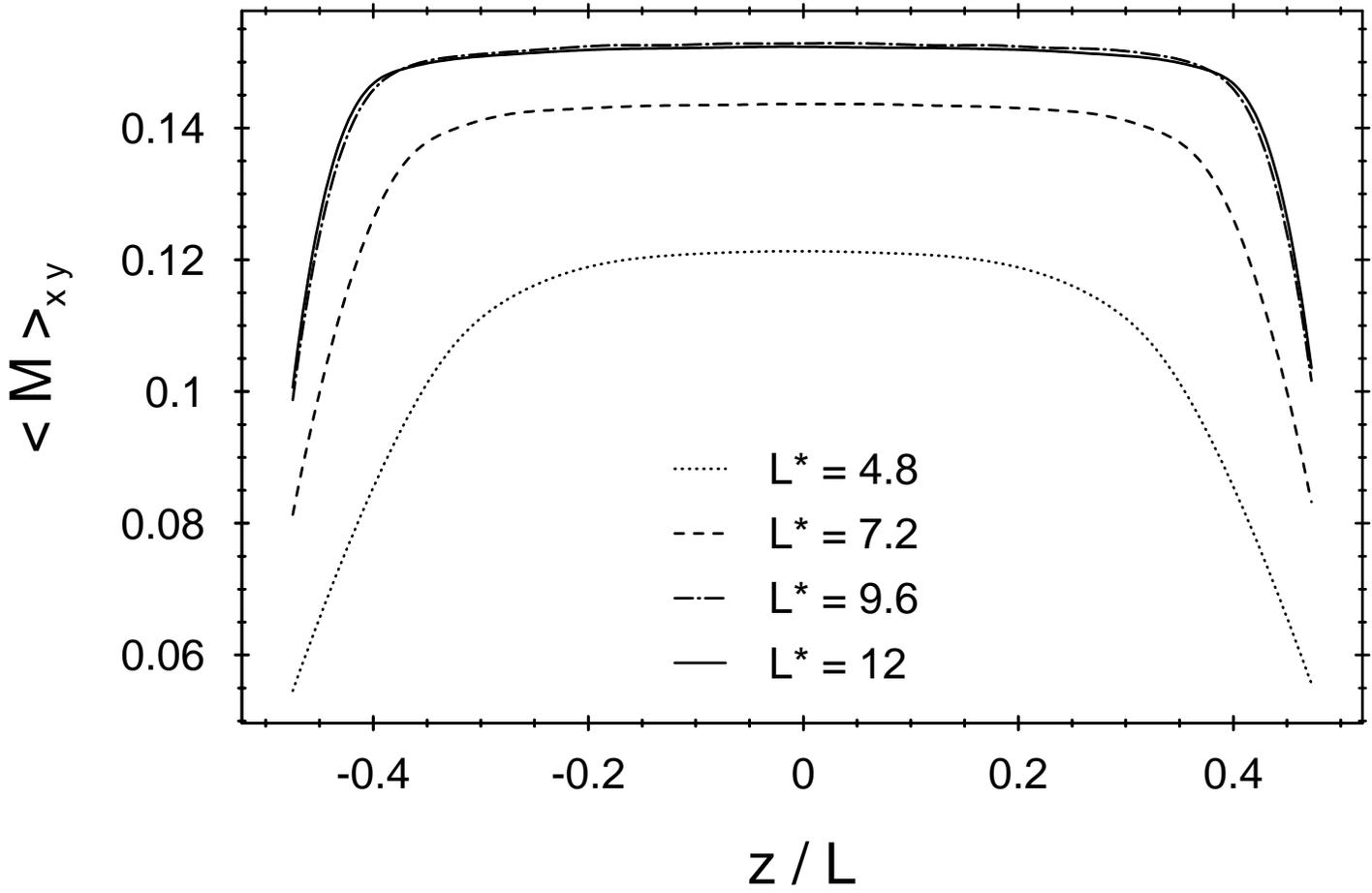
\begin{figure}
\caption{Averaged absolute value $\langle M \rangle_{xy}$ of the magnetization within the planes
$z=\text{const}$ for $N=20$ and several system sizes. It is constant
in the inner part of the sample and decreases near the bottom and top
surfaces. The size of the disordered region remains proportional to
$L$ for large $L$.}
\label{figmbetrz}
\end{figure}

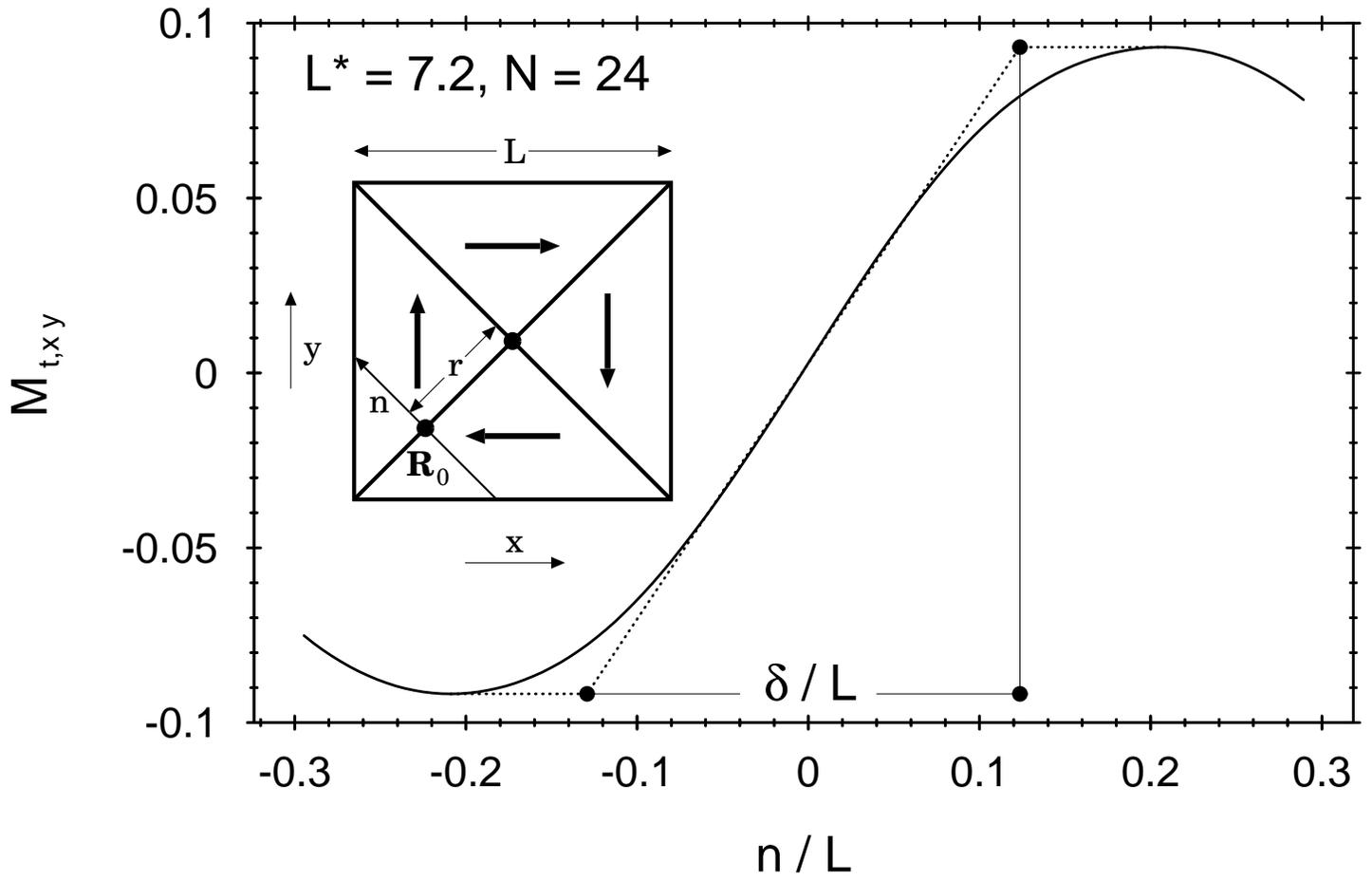
\begin{figure}
\caption{Variation of the transversal magnetization component
$M_{t,xy}=(M_1+M_2)/\protect\sqrt{2}$ upon traversing the domain wall
along a normal to the wall through the point
$\Rv_0/\sigma=(-1.95,-1.95,-0.15)$ (see the inset) for $\ls=7.2$ and
$N=24$ (compare \abbpref{figschichtz}(a)). This is the component which characterizes the difference of
the magnetization directions of the adjacent domains. Our definition
of the wall thickness $\delta$ can be inferred from the figure. Here $r/L=0.38$.}
\label{figmtxy}
\end{figure}

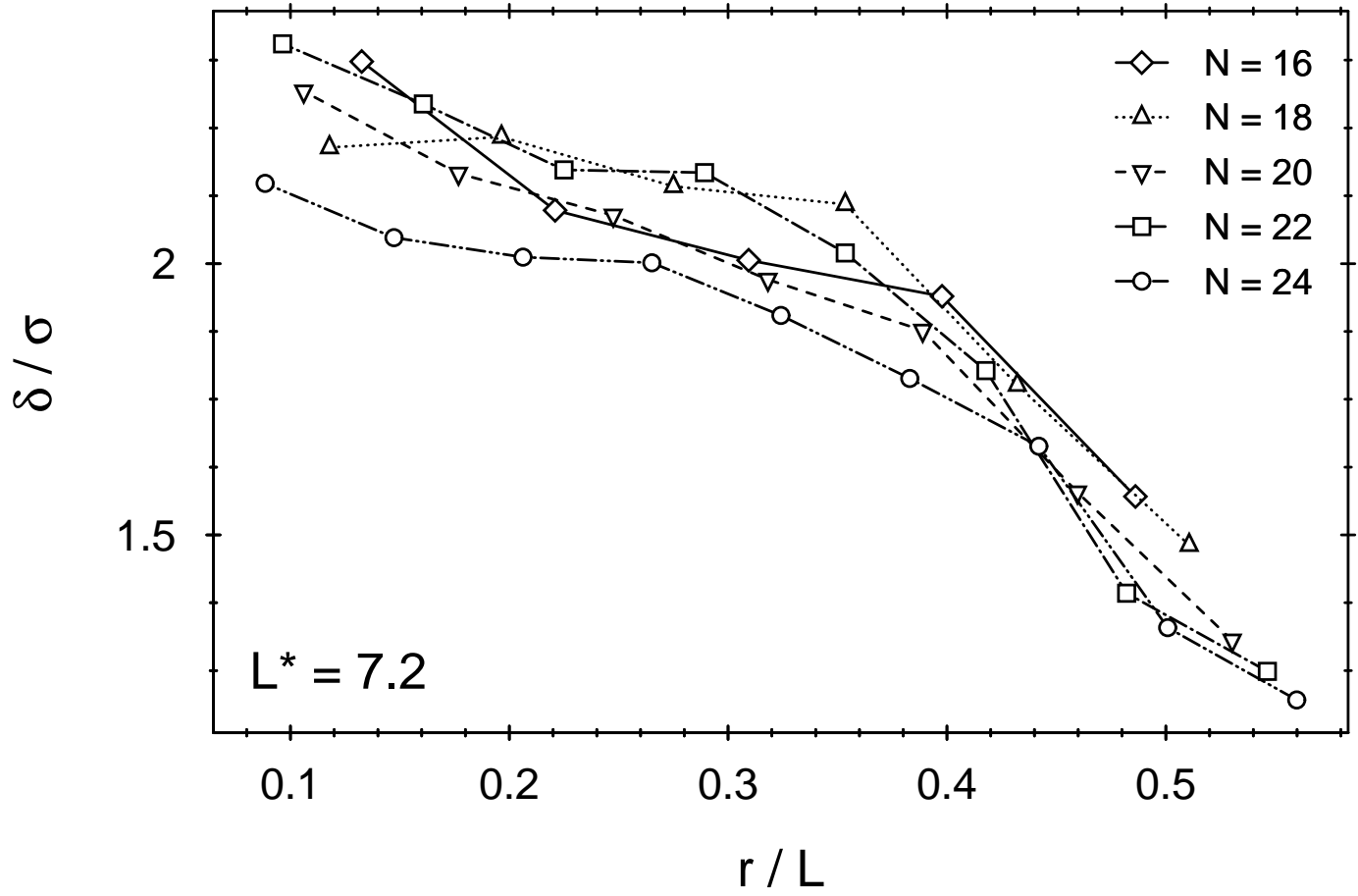
\begin{figure}
\caption{Dependence of the domain wall thickness $\delta$ on the
distance $r$ from the center of the plane $z=-a/2\simeq 0$ (i.e.,
close to the midplane) for $\ls=7.2$
and different numbers of lattice points $N$, averaged over the four
equal walls in the cube (see \abbpref{figmtxy}). In the medium range of $r$ the thickness
$\delta$ depends only weakly on $r$ and $N$, while near the center and
edges $\delta$ is influenced by the escape of the orientation into the
$z$ direction and the vicinity of the surfaces, respectively.}
\label{figthick}
\end{figure}

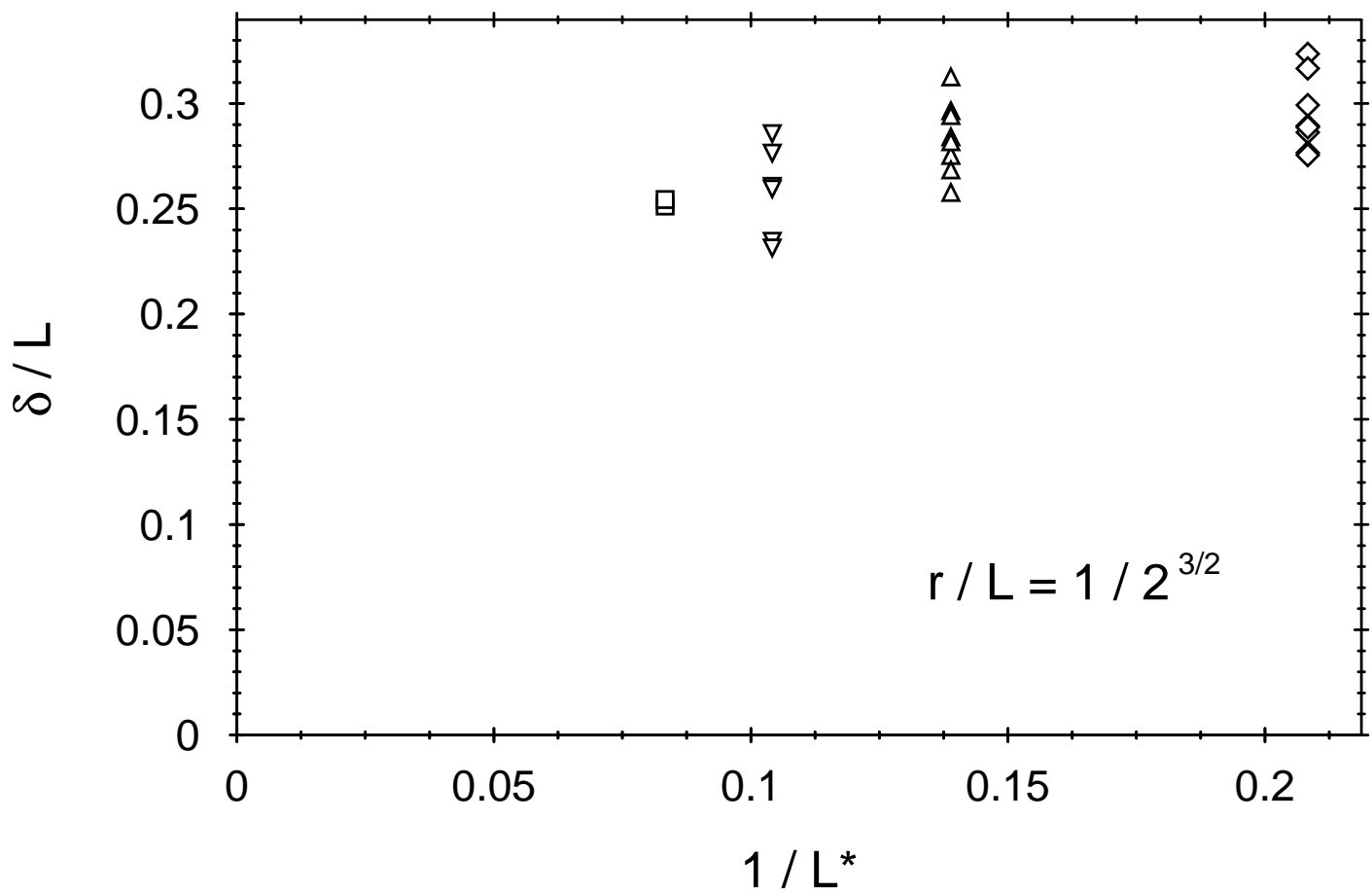
\begin{figure}
\caption{The wall thickness at half distance between the center and
the edges of the cube as a function of the system size $L$. The
different points at the same $L$ correspond to different lattice
constants. (From top to bottom: $\ls=4.8$,
$N=10,12,14,16,20,18,22,24$; $\ls=7.2$, $N=14,12,18,10,22,16,20,24$;
$\ls=9.6$, $N=18,14,16,24,20,22$; $\ls=12$, $N=22,20$.) The extrapolation to $1/L=0$ suggests a finite value of
$\delta/L$ in the thermodynamic limit.}
\label{figmidthick}
\end{figure}

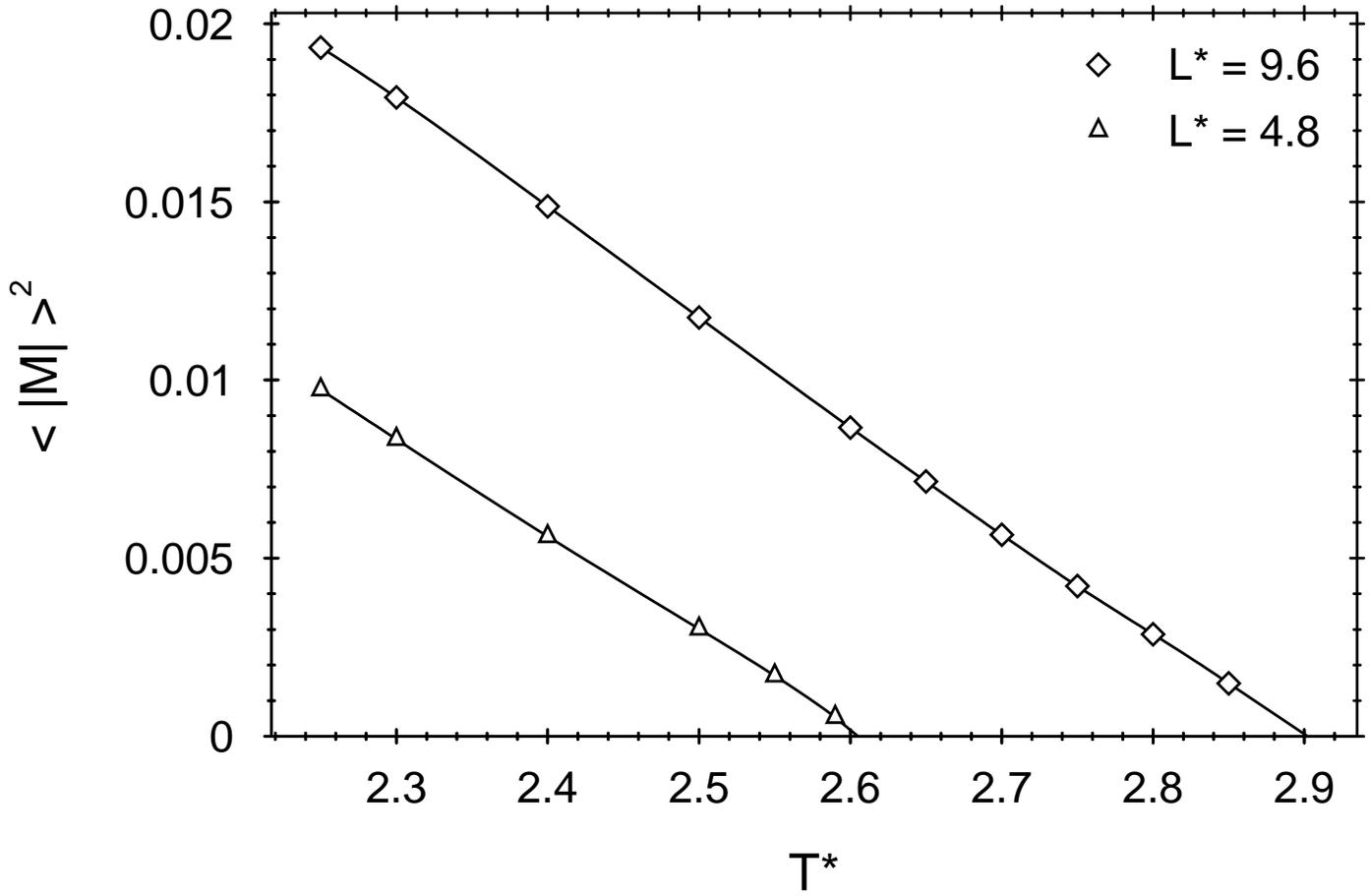
\begin{figure}
\caption{The squared spatially averaged order parameter $\langle |\Mvec|
\rangle^2=\left(N^{-3} \sum_\Rv |\Mvec(\Rv)|\right)^2$ decreases
linearly with increasing temperature $\ts$. It vanishes at a critical
temperature $T_c(L)$ which depends on the system size. This linear
behavior holds even outside the close vicinity of $T_c(L)$ where it
has to be so due the inherent mean-field character of the present
theory.}
\label{figmagnt}
\end{figure}

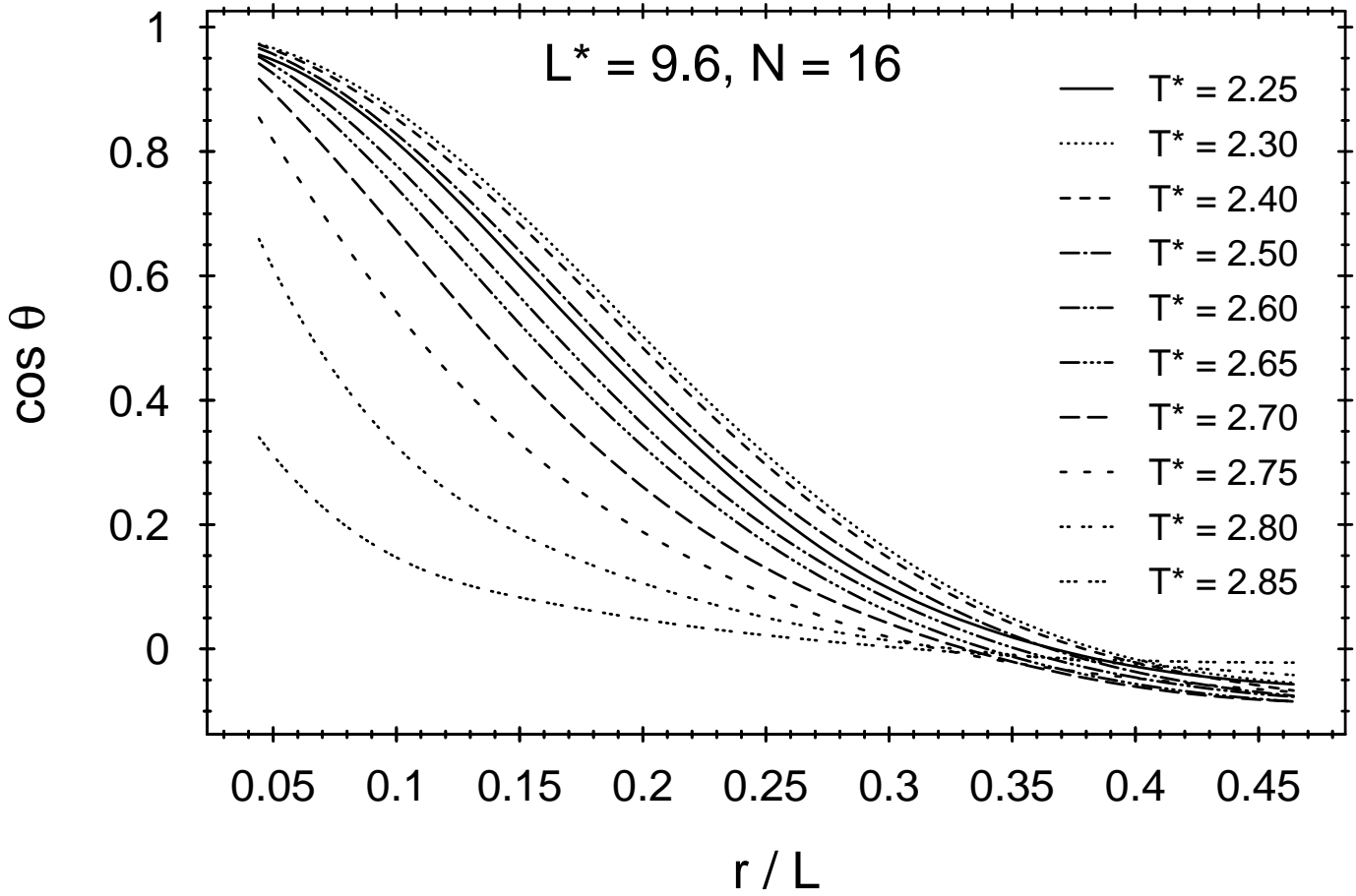
\begin{figure}
\caption{The polar angle of $\Mvec$ (see Figs.~\pref{figschichty}(c)
and \pref{figauswmd}) along
the central lines near the midplane (fixed $z=-a/2$ as in
\abbpref{figauswmd}) as a function of the temperature for $\ls=9.6$ and
$N=16$. The escape near the vortex axis reduces upon approaching the
critical point $\ts_c(\ls=9.6)=2.89$.}
\label{figauswmt}
\end{figure}

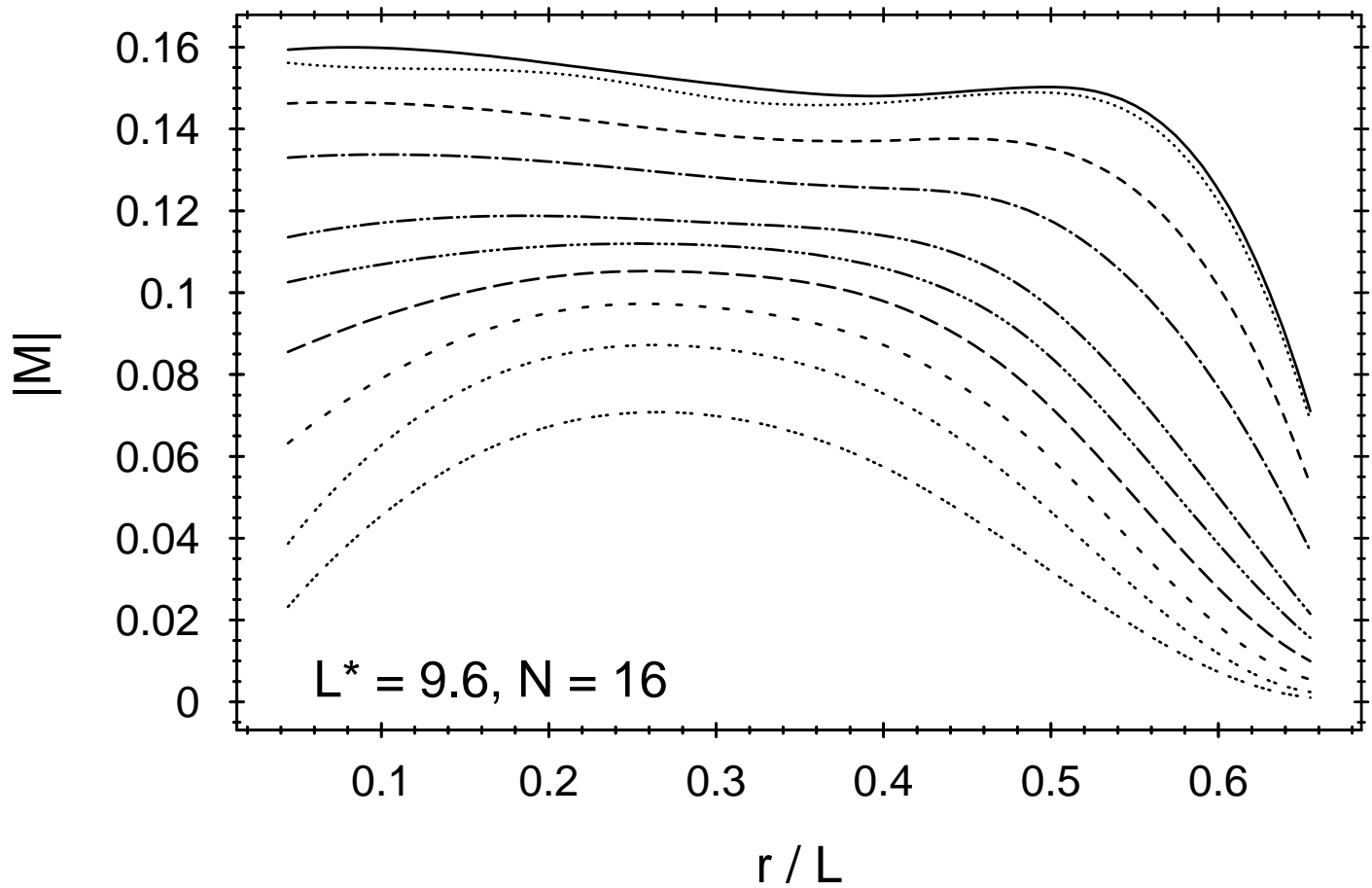
\begin{figure}
\caption{The absolute value of the magnetization along the diagonal
lines (for fixed $z=-a/2$) at various temperatures. The different line styles
correspond to the same temperatures as  in \abbpref{figauswmt}. Near the
phase transition $\ts_c(\ls=9.6)=2.89$ the orientational order is
particularly reduced near the vortex
axis and in a surface layer of increasing thickness.}
\label{figmbetrdt}
\end{figure}

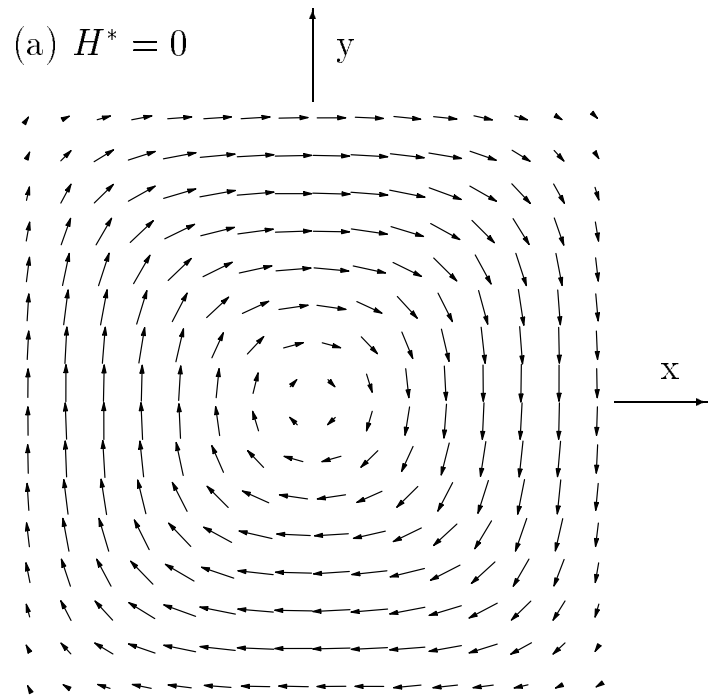
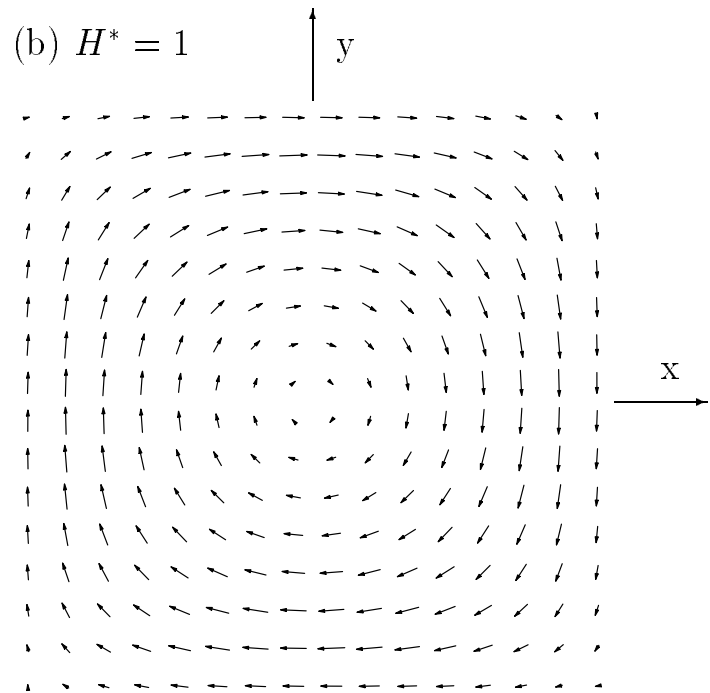
\begin{figure}
\caption{Sections orthogonal to the field direction through the magnetization structure for
$\ls=9.6$ and $N=16$ at $z^\ast=-0.9$ (a) without and (b) with an
external field $H^\ast=H \protect\sqrt{\sigma^3/\epsilon}=1$ applied
in  the $z$
direction. The scale factor which determines the lengths of the arrows
is the same  in both parts of this figure and also in
\abbpref{figschichtyh}. The transversal components of the
magnetization are reduced by the
field but the overall feature of the inhomogeneous structure is preserved.}
\label{figschichtzh}
\end{figure}

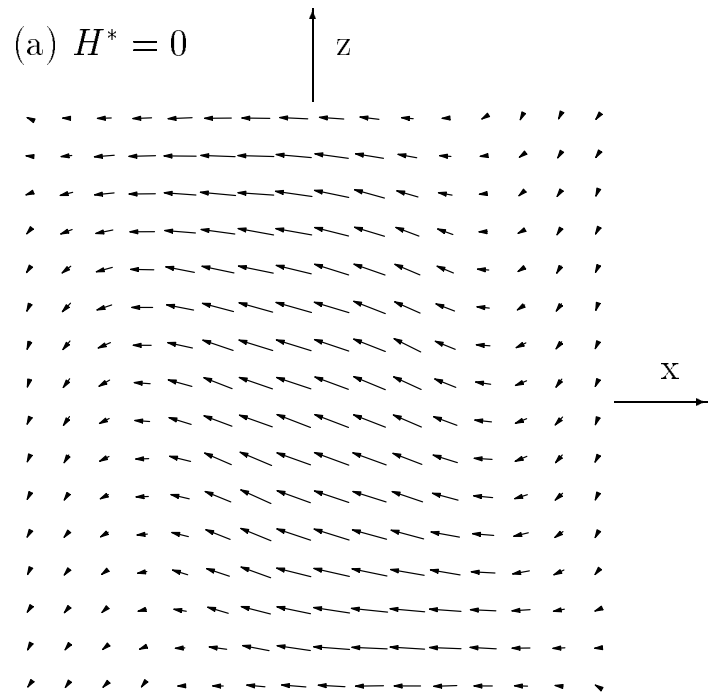
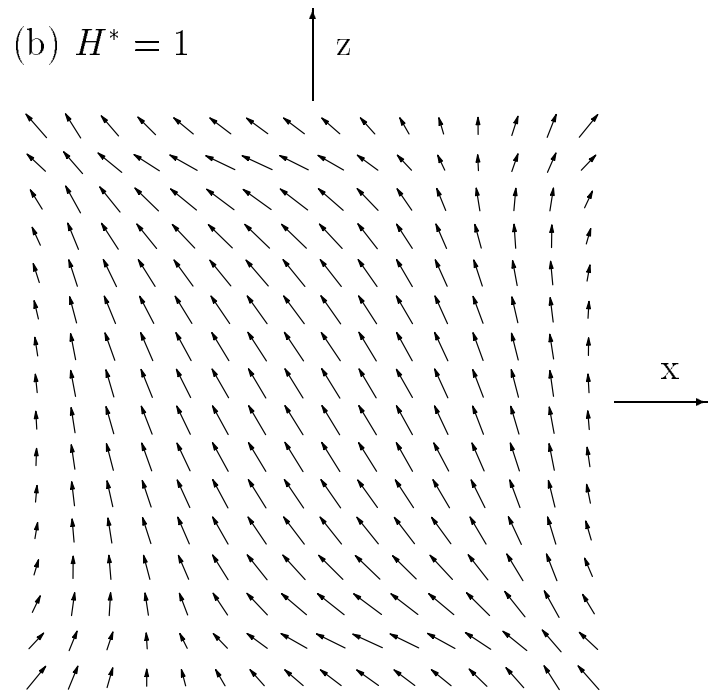
\begin{figure}
\caption{Vertical sections perpendicular to the $y$ axis and thus
parallel to the field direction through the
same magnetization structures as in \abbpref{figschichtzh} for
$y^\ast=-2.1$. At this distance of the plane from the center the
$z$ component of the magnetization in zero field is only small, while the external field induces a large
$z$ component everywhere.}
\label{figschichtyh}
\end{figure}

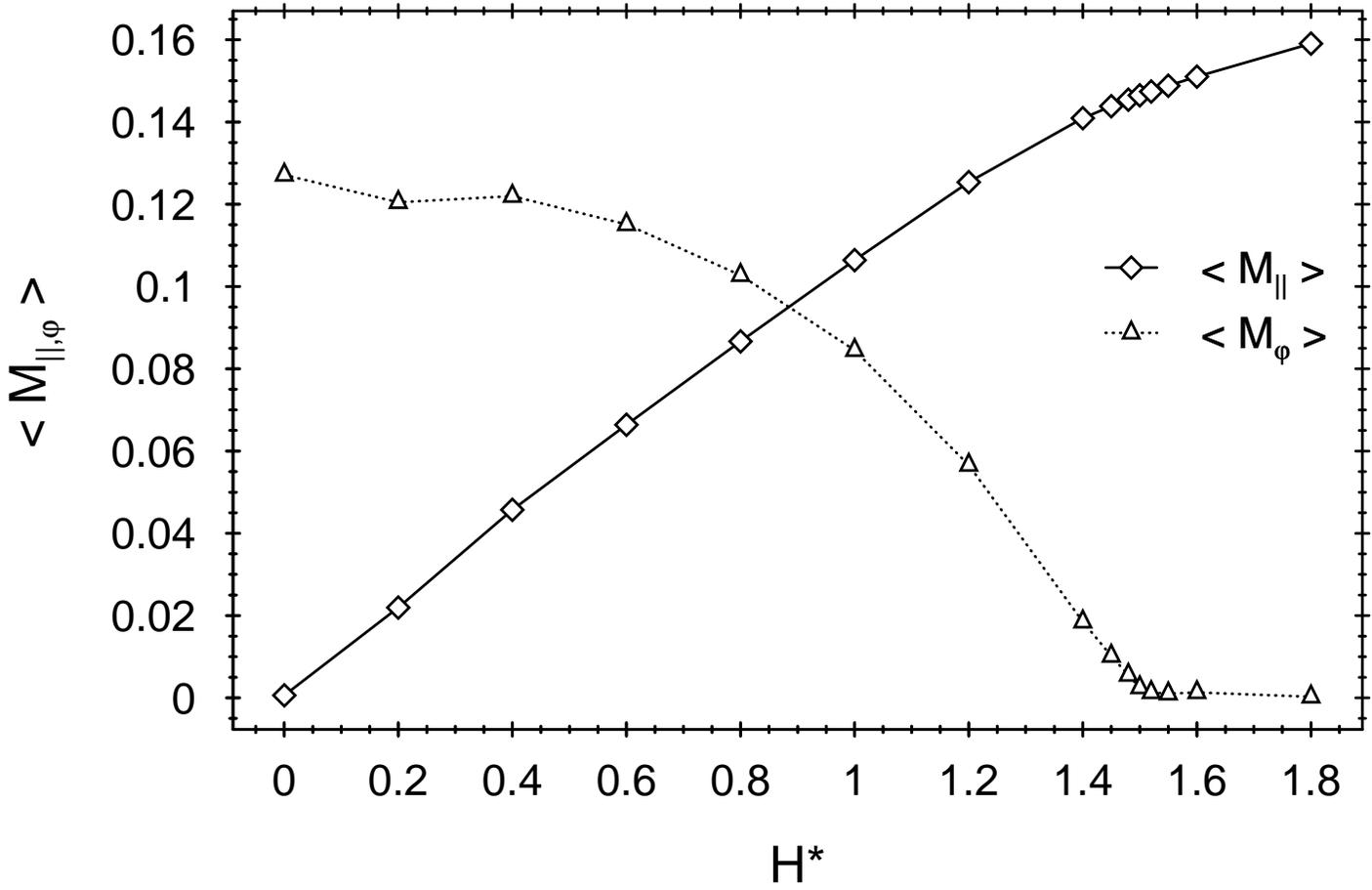
\begin{figure}
\caption{Dependence of the averaged longitudinal magnetization
component $\langle M_\parallel\rangle$ and the averaged angular
component $\langle M_\varphi\rangle$ on the field strength ($\ls=9.6$,
$N=16$, $\ts=2.25$). The angular component decreases for increasing $H^\ast$ and
vanishes seemingly linearly at $H_c^\ast\simeq 1.50$. Below $H_c^\ast$ the parallel component increases
approximately linearly and crosses over towards saturation for
$H^\ast>H_c^\ast$. Above $H_c^\ast$ the sample has an approximately
homogeneous magnetization. Compare Fig.~11 in
Ref.~\protect\cite{Domains}.}
\label{figmagn}
\end{figure}

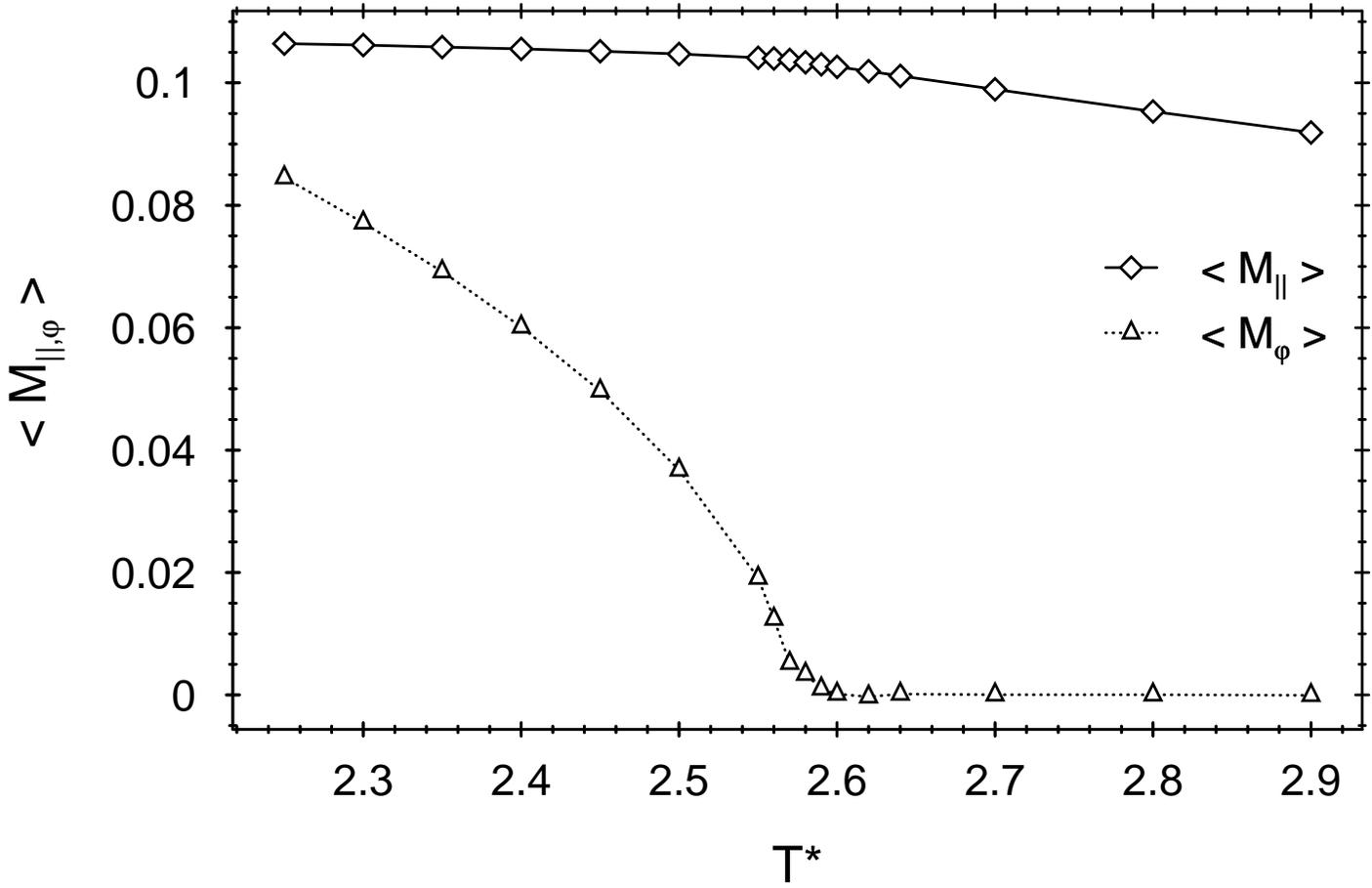
\begin{figure}
\caption{The same magnetization components as in \abbpref{figmagn} for
  fixed $H^\ast=1$ as a function of temperature ($\ls=9.6$, $N=16$). There
  is a transition from a vortex structure to an almost homogeneous
  magnetization at $\ts_c(H^\ast)\simeq 2.65$. Compare Fig.~12 in
Ref.~\protect\cite{Domains}.}
\label{figmagnhfix}
\end{figure}

Based on density-functional theory we have obtained the following main
results for the long-ranged orientational order of a dipolar liquid
confined to a cube:

\begin{enumerate}

\item The equilibrium magnetization configuration corresponds to a
predominantly planar vortex
structure which can be viewed as being composed of four triangular domains
separated by thick domain walls along the diagonal planes of the cube. Near
 one of the symmetry axes of the cube chosen spontaneously the
magnetization escapes into the third dimension in order to
avoid a topological line singularity (Figs.~\ref{figschichtz},
\ref{figschichty}, and \ref{figauswmd}).

\item Point defects arise at the centers of the top and bottom face of
the cube. The absolute value $|\Mvec|$ of the magnetization is reduced
inside the cores of these defects (Fig.~\ref{figmpdef}) whose sizes
scale with the system size. Therefore an appropiate description of the 
structure requires to take the spatial variation of $|\Mvec|$ into account.

\item Near the surfaces there is a layer of reduced orientational order
(Figs.~\ref{figmbetrmd} and \ref{figmbetrz}).

\item The domain walls (\abbref{figmtxy}) are mainly of the N\'eel type, i.e., upon
traversing the wall the magnetization rotates within the plane spanned
by its asymptotic directions deep inside the adjacent domains.
 The size dependence of the wall
thickness (\abbref{figmidthick}) and of other features of the configurations are compatible
with the scaling behavior formulated in \eqref{scale}.

\item The ferromagnetic order vanishes at a critical temperature which
depends on the system size (\abbref{figmagnt}). Upon increasing the
temperature the ``escape'' region near the vortex axis is gradually
replaced by a column of disordered fluid (Figs.~\ref{figauswmt} and
\ref{figmbetrdt}).

\item In weak external fields normal to the surfaces of the cube the
  vortex axis is aligned parallel to the field direction and the
  overall structure is similar to the one in zero field
  (Figs.~\ref{figschichtzh} and \ref{figschichtyh}). For stronger
  fields the magnetization is rotated increasingly into the field
  direction until a transition to an approximately homogeneously
  magnetized state takes place at a critical field strength beyond
  which the vortex structure is lost (\abbref{figmagn}). In an
  external field the magnetization components normal to the field
  direction vanish linearly upon approaching a critical temperature
  (\abbref{figmagnhfix}).

\end{enumerate}

Recently metallic liquid ferromagnets have been discovered in
undercooled CoPd alloys \cite{Platzek,Reske,Albrecht}. For the formation of the
ferromagnetic
order of these materials short-ranged exchange interactions play an
important role. However, in addition dipolar interactions are also present
and, as in solid ferromagnets, are essential in forming
the domain structure. Therefore for these systems we expect strong similarities to
the behavior of dipolar fluids as discussed in this work. Up to now these
undercooled liquid alloys have been prepared only as electromagnetically suspended
spherical samples. One can speculate that the domain structure within
a sphere has a vortex axis, too, but no domain walls due to the higher
symmetry compared to a cube. Our analysis indicates that an
experimental study of these structures (e.g., by using magnetic
neutron or X-ray tomography \cite{XMCD}) would certainly be very rewarding.

\acknowledgements 

One of us (S.D.) acknowledges helpful discussions with D.~Thouless.


\end{document}